\documentstyle[12pt]{article}
\newcommand{\bq}{\begin{equation}}
\newcommand{\ee}{\end{equation}}
\newcommand{\arr}{\rightarrow}
\newcommand{\fr}[2]{\frac{#1}{#2}}

\newcommand{\al}{\alpha}
\newcommand{\De}{\Delta}
\newcommand{\defi}{\De \fii}
\newcommand{\eps}{\varepsilon}
\newcommand{\fii}{\varphi}
\newcommand{\fin}{\varphi_0}
\newcommand{\gm}{\gamma}

\newcommand{\so}{\vec{\sigma}_1}
\newcommand{\sd}{\vec{\sigma}_2}
\newcommand{\sisi}{\vec{\sigma}_1 \vec{\sigma}_2 }

\newcommand{\aav}{{\bf a}}
\newcommand{\bv}{{\bf b}}
\newcommand{\kv}{{\bf k}}
\newcommand{\pv}{{\bf p}}
\newcommand{\qv}{{\bf q}}

\begin{document}

\vspace*{1.5cm}

\begin{center}
{\large
The "recoil" correction \mbox{$\sim m \al^6$} to hyperfine
splitting of positronium ground state.  }

\vspace{1.0cm}
A.P.Burichenko 

\end{center}

\vspace{1.0cm}
    
\begin {abstract}

The "recoil" correction of order \mbox{$m \alpha^6$}  to the hyperfine
splitting of positronium ground state was found. The formalism employed
is based on the noncovariant perturbation theory in QED.
Equation for two-particle component of full (many-body) wave function
is used, in which effective Hamiltonian depends on the energy of a
system. The effective Hamiltonian is not restricted to the
nonrelativistic region, so there is no need in any regularization.
To evaluate integrals over loop momenta,  
they are divided into "hard" and "soft" parts, coming from large and
small momenta respectively. 
Soft contributions were found analytically,
and hard ones are evaluated by numerical integration.
Some soft terms due to the retardation cancel each other. 
To calculate the "hard" contributions, a great number of noncovariant
graphs is replaced by only a few covariant ones.
The hard contribution was found in two ways.
The first way is to evaluate contributions of separate graphs,
using the Coulomb gauge. The second one is to calculate
full hard contribution as a whole using the Feynman gauge. 
The final result for the "recoil" correction is 
\mbox{$0.381(6) m\al^6$}  
and agrees with those of \cite{Pa,Elh}. Diagram-to-diagram comparison 
with the revised results of \cite{AS} was done. All the results  
agree, so the "recoil" correction is now firmly established. This
means a considerable disagreement with the experimental data. 

\end{abstract} 
\vspace*{1.5cm}

    1. INTRODUCTION.

\vspace{0.5cm}

    The treatment of relativistic bound states is one of challenging
tasks of Quantum Electrodinamics. Positronium, the bound state of
the electron  and the positron, is one of the most appropriate objects for
theoretical and experimental study of relativistic bound states.
Because of the small masses of its constituents, effects of the strong and
the weak interaction are negligible compared with the accuracy of current
experiments on the positronium spectroscopy. On the other hand this
experimental  accuracy is good enough for comparison with results of
modern theoretical investigations, which reached accuracy level
\mbox{$\sim m \al^6$} for contributions to the energy. Therefore, now
positronium spectrum can be investigated within QED framework.

Now the most accurately measured positronium property is hyperfine
splitting of the ground state, i.e. energy difference between the
$1^3S_1$ and $1^1S_0$-states, denoted below as
 $\Delta \nu$. The two best experimental results for this value are
\bq\label{exp1}
 \Delta \nu\ =\ 203\ 387.5(1.6)\ MHz\ \; ,  \ee
\bq\label{exp2}
 \Delta \nu\ =\ 203\ 389.10(0.74)\ MHz\ \; ;  \ee
obtained in \cite{Ex1,Ex2} and \cite{Ex3} respectively.

Calculation of $\Delta \nu$ has a long history; contributions
\mbox{$\sim m \al^4$},
\mbox{$\sim m \al^5$}, \mbox{$\sim m \al^6 \ln{\al}\ $} were found
in \cite{qm1,qm2,KK,FM,L77,BY}; their sum equals 
\bq\label{denu456}  m \al^4\
\left[ \; \fr{7}{12}\ -\ \fr{\al}{\pi}\ \left(
 \; \fr{8}{9}+\fr{1}{2} \ln{2} \right) \;
-\ \fr{5}{24}\ \al^2 \ln{\al}\ \right] \; =\
203\ 400.29 MHz\ \; .  \ee
To compare experimental results (\ref{exp1},\ref{exp2}) with the
theoretical one, the contributions \mbox{$ \sim m \al^6 $} (without the
logarithm) are to be obtained. An essential progress in their
calculation was made only recently.

 There are several sets of corrections $\sim m \al^6$ to the hyperfine
splitting, that are different in origin and
may be calculated independently. First, these are three
sets consisting of contributions arising from the one-, two-, and
three-photon annihilation; they were found in
\cite{AdAn1,HLZ}, \cite{AdAn2} and \cite{AdAn3,DeR} respectively.
Second, there are contributions associated with radiative corrections
to the Breit potential, i.e. those of formal order
$(Z \al)^4\ \al^2 m $), found in \cite{BE,BMR}.
Third, there exist "recoil" corrections, i.e. those coming from the graphs
in which all photon lines connect two fermion ones; these corrections
are of conventional order \mbox{$ (Z \al)^n\ m $} (here $Ze$ is the charge
of one of the particles;  \mbox{$ Z=1$} in positronium).
Finally, there are radiative-recoil
(i.e. $ \sim (Z \al)^5\ \al m $) corrections found in
 \cite{RR1,RR2,Elh1}.

Below the sum of contributions
\mbox{$ \sim m (Z \al)^6,\ \; \sim m (Z \al)^6 \ln{\al}$}
to \mbox{$ \De \nu$} is denoted as \mbox{$ \De \nu_{rec}$}.
Calculation of its nonlogarithmic part is the most difficult step of
calculation of \mbox{$ \De \nu$}. This was the subject of works
 \cite{CL86,Pa,AS,Elh}. The present work is also devoted to calculation of
\mbox{$ \De \nu_{rec}$}.

The results of the three first works, namely \cite{CL86,Pa,AS}, were all
different.
The result of work \cite{Elh} 
coincides essentially with the result of \cite{Pa} (they are equal to
\mbox{$ 0.3763 m \al^6 $} and \mbox{$ 0.3767(17) m \al^6 $}
respectively). The result of the present work is
\mbox{$ 0.381(6) m \al^6 $} and also coincides, within its accuracy,
with the results of \cite{Pa,Elh}. 
Recently the result of \cite{AS} was corrected by its authors. 
Now it is equal to \mbox{$ 0.3764(35) m \al^6 $}
and agrees with the results mentioned above.

In \cite{CL86} the so--called NRQED (nonrelativistic QED),
an effective field theory equivalent to QED,
was formulated and then applied to calculation of
\mbox{$ \De \nu_{rec}$}; nonlogarithmic part of the result equals 
\mbox{$  0.167(33) m \al^6 $}. However, just now the preliminary  
result was obtained in the NRQED framework \cite{HL}, which is
consistent with
that of \cite{Pa,Elh}. In \cite{AS} the calculation was performed by
Bethe-Salpeter formalism. 
In \cite{Pa} the calculation was done by an effective Hamiltonian approach;
the method employed in \cite{Elh} in fact also uses an effective
Hamiltonian. Difference between \cite{Pa} and \cite{Elh} is in
different regularizations used; besides that, the result of \cite{Pa} was
obtained by numerical integration, whereas that of \cite{Elh} is
analytic.  The effective Hamiltonian approach is essentially
a combination of ideas of the NRQED and the old--fashioned noncovariant
perturbation theory in
QED. The latter was used first for QED bound state calculations in
 \cite{KMY}.

The present paper also employes a formalism based on old-fashioned
noncovariant ("time-ordered") perturbation theory for QED. Starting
from the Shr\"{o}dinger equation for full (many-body) wave function one
easily obtains an equation containing only two-particle part of the wave
function, in which effective Hamiltonian depends on the energy of a
system. Then the effective Hamiltonian is divided into unperturbed
part and perturbation,
zeroth approximation for the two-particle wave function is found, and
using them
corrections to the energy levels are calculated by means of
usual Rayleigh--Shr\"{o}dinger perturbation
theory. This calculation consist essentially in finding of expectation
values of operators corresponding to graphs of the noncovariant
perturbation theory over the
"unperturbed" wave function. To evaluate the integrals over loop
momenta each of
the integrands is divided into two parts, "hard" and "soft"; here terms "hard"
and "soft" mean that in the desired order a "hard" part entirely arises from
the region of momenta \mbox{$\sim m$}, whereas for a "soft" part the
region of momenta
 \mbox{$\sim \al m$} is also essential. This decomposition is
performed in such a way that "soft" contributions may be easily found
analytically, and "hard" ones are found by numerical integration.
To calculate the "hard" contributions with the required accuracy 
external legs of the
graphs may be set on the mass shell, which allows to replace the sum of
 a great number of noncovariant graphs by the sum of
only a few covariant ones.

The method of calculation described above differs from that of
\cite{Pa} and \cite{Elh} in the way effective Hamiltonian is defined. In
\cite{Pa} and \cite{Elh}, effective Hamiltonian is constructed so as
to reproduce the
scattering amplitudes, whereas in the present paper it is immediately
derived from the full QED Hamiltonian. Besides that,
 the effective Hamiltonian used in the present
work is not restricted to the nonrelativistic region, in difference with
\cite{Pa} and \cite{Elh}, so there is no need in regularization of
the effective Hamiltonian or matrix elements it enters.

The plan of the rest of the paper is following. In section 2, I
briefly describe the formalism used in the paper, i.e. the effective
"Shr\"{o}dinger equation" for two-particle component of the wave function,
"zeroth"
approximation to the wave function, perturbation theory formulas
for the calculation
of \mbox{$\De \nu$}; these formulas are transformed into form of
expectation values over the nonrelativistic Coulomb wave function.
In section 3, I describe the general method of calculation of these expectation
values, namely, the explicit manner to divide them into "soft" and "hard"
parts; correspondence between 
 these values and covariant graphs (and also
scattering amplitudes on mass shell) is established. In section 4
this general method is applied to contributions of various sets of graphs in
turn. Section 5 contains the description of the procedure of numerical
evaluation of "hard" contributions, the checking of this procedure,
and comparison of the
results obtained with the results of \cite{Pa,Elh,AS}. Section 6 consist
of summary including comparison with experimental results.

\vspace{1.5cm}

2. FORMALISM.

\vspace{0.5cm}

The starting point of the formalism used in the present paper is the
Shr\"{o}dinger equation
\bq\label{eq:uSh}
 H | \psi_N > = E  | \psi_N >\ \; ,  \ee
where the positronium wave function \mbox{$ | \psi_N > $}
includes components with different number of particles.
In (\ref{eq:uSh})  $H$ is the full QED Hamiltonian,
 \mbox{$ H = H_0 +V_N\ ,\ \; $}
where $H_0$  is the free Hamiltonian, $V_N$ is the
interaction.

Direct use of the many-body wave function is unconvenient.
However in the lowest order in $\al$ the only nonvanishing component of
\mbox{$ | \psi_N >$} is two-particle one, which is denoted as
 \mbox{$ | e >$}.
It is easy to obtain an equation containing only \mbox{$ | e >$}.
   Let   $P_e$ be projector onto the two-particle subspace,
 \mbox{$P_a = 1 - P_e$},
\[  | e >\ \equiv \; P_e  | \psi_N >\  , \; \; \;
 | a >  \; \equiv \; P_a  | \psi_N > \ \; . \]
From the Shr\"{o}dinger equation one gets
\bq\label{eq:u1}
 P_e H P_e  | e >\ +\ P_e H P_a  | a >\ =\ E |e>\
  \; ,  \ee
\bq\label{eq:u2}
 P_a H P_e  | e >\ +\ P_a H P_a  | a >\ =\ E |a>\
  \; .  \ee
 Expressing \mbox{$ | a >$} from (\ref{eq:u2}) and substituting
 it into (\ref{eq:u1}) results in
\bq\label{eq:ue}
(\ P_e H P_e\ +\ V_a (E)\ -\  E\ )\ | e >\ =\ 0\ \; ,  \ee
       where
\[  V_a (E)\ =\ P_e H P_a\ \frac{1}{(E - P_a H P_a)}\
P_a H P_e\  \;  .  \]
  Energy levels which could be obtained from (\ref{eq:ue})
are derived in a form of perturbative expansion.
For this purpose "Hamiltonian" \mbox{$ P_e H P_e\ +\ V_a (E)\ $} is
divided into the unperturbed part
\mbox{$ H^{(0)}\ \equiv\ P_e H_0 P_e + V_0\ \; $}
and the perturbation
\mbox{$ V(E) \equiv\ V_a(E) + P_e V_N P_e - V_0\ , \; $}
where $V_0$ is the usual nonrelativistic Coulomb interaction.
Contributions to the energy, which are of 1-st, 2-nd, and 3-rd order
in $V$, 
have leading orders (besides logarithms)
$ m \al^4,\ m \al^5 $ and $ m \al^6 $ respectively.

 Clearly
 \mbox{$ V(E)\ +\ V_0\ =\ V_a(E)\ + P_e V_N P_e\ $}  is
the two-particle irreducible kernel for the Green function appearing in
the noncovariant perturbation theory (for transitions between
two-particle states). Its
expansion in $\al$ is
 \[ V(E)\ +\ V_0\ =\ P_e V_N\ \left( \; 1\ +\
 \fr{P_a}{E-H_0} V_N\
 +\ \fr{P_a}{E-H_0}\ V_N\ \fr{P_a}{E-H_0}\ V_N\ +\ .........\
 \right) \; P_e\ \; \; \;   \]
and corresponds to the set of two-particle irreducible graphs of the
noncovariant
perturbation theory (here and below "irreducible" is meant in
noncovariant sense, for
instance, graph  shown in fig.1 is irreducible).

Since diagram technique for the noncovariant theory is not of common
use, its rules
are described here. External fermion lines correspond to
positive-energy spinors normalized as \mbox{$u^{+} u = 1 $} (for
convenience, the fermion-fermion channel, instead of the fermion-antifermion
one, is considered). Internal fermion lines 
are described by the projectors
\[  \Lambda ^{\pm} (\pv)
= ( \eps_p \pm ( \gm_0 + \al \pv ) ) / \; (2 \eps_p)\ \; ,
\; \; \; \; \; \eps_p \equiv \sqrt{\pv^2+m^2}\ \; ;  \]
factors corresponding to external lines may be written in form
\[ u (\pv) \; = \; ( (2 \eps_p) /(\eps_p + m) )^{1/2}\ 
\Lambda^{+} ( \pv )
\left( \begin{array}{c}  1 \\ 0 \end{array} \right)  w\ \; ,
 \]
\mbox{$\left( \begin{array}{c}  1 \\ 0 \end{array} \right) $}
is
\mbox{$( 2 \times 4 )  $} matrix, \mbox{$ \; \; w  $} is a two-spinor
normalized as \mbox{$ \; w^{+} w = 1\ . $}
The two fermions are assumed to have charges of opposite sign; then
Coulomb photon with the momentum $\qv$ gives the factor
\mbox{$ \; -  4  \pi  \al / \qv^{2} \; \; , \; \;$}
and magnetic (transverse) photon gives the factor
\mbox{$ \; (\ - 4 \pi \al\ )\
(\ \vec{\al}_1\vec{\al}_2\ -\ (\vec{\al}_1 \qv)\ (\vec{\al}_2 \qv)\
/q^2\ )\ /2q\ ,\ $} (in the noncovariant perturbation theory energy
of virtual photon
is equal to its momentum). Every negative-energy projector
$ \; \Lambda^{-}\ $ gives extra factor $(-1)$. Factor
\mbox{$ \; (\ E -  E_k + i0\ )^{-1}$} corresponds to each
intermediate state, where $E_k$ is energy of the intermediate state,
$E$ is full energy of a system. Loop momenta $ \; \qv_i$ which remain
undetermined lead to the integration
 \mbox{$  \; d^3 q_i\ /\ (2 \pi)^3\ $}.

In the rest of the paper all operators and wave functions refer to
the two-particle subspace. All the calculations are performed in the
center-of-mass frame, so the only variable except spins is $\pv$,
i.e. momentum of particle 1 in the c.m.f.. Eigenstates of $\pv$ are
normalized as usual, according to
\mbox{$ < \pv' | \pv >\ =\ (2 \pi)^3\ \delta^3 ( \pv - \pv' )\ , $}
kernel of an arbitrary operator $X$ is denoted as
\mbox{$ X (\pv, \pv')\ \equiv\ < \pv' | X | \pv > $}.

 The kernel  \mbox{$ V (\pv,\pv';\ E)\ +\ V_0 (\pv,\pv')\  $}
is given by the sum of matrix elements corresponding to all the
irreducible graphs for initial and final states having momenta
$\pv$ and $\pv'$ respectively.
The tree diagram with Coulomb photon corresponds to operator $V_c$, and
that with magnetic photon corresponds to $V_m$.
Let us write $V$ in the form
 \bq\label{eq:V}
  V (E)\ =\ V_1\ +\ V_2\ +\ V_3\ +\ ......\ \; \; \; ,  \ee
where
\[ V_1\ =\ V_{1c}\ +\ V_m\ \; \; ,\ \; \; \;
V_{1c}\ \equiv\ V_c\ -\ V_0\  \; ,  \]
 $ V_2,\ V_3\ $ and so on correspond to sums of the irreducible graphs
having two, three, and more, photon lines respectively. Being
expanded in $ v/c $ up to the second order,
 \mbox{$ \; H^{(0)} + V_1\ $} gives the Breit Hamiltonian.

Consider, for example, the first-order in $V$ correction to energy
levels. The first term in (\ref{eq:V}) gives to this correction
contribution of formal order
 \mbox{$ \; \sim m \al^4\ $}, the second term gives
 \mbox{$ \sim m \al^5\ $}, and so on.
 However in fact this expansion does not converge: for instance,
spin-independent correction to the energy, arising from
$V_2,\ V_3\ $ and further terms of expansion (\ref{eq:V}), are all of
order \mbox{$ \sim m \al^5\ $}. This takes place due to graphs
shown in fig. 2: (a), (b), (c), and so on, because these graphs produce
"ultrasoft" contributions, i.e. those coming from the region where
momentum of the virtual magnetic photon is \mbox{$ \sim m \al^2\ $}.
Fortunately for the hyperfine splitting the "rest term" of this expansion
corresponds to the graphs first of which is graph shown at fig 2(c), and has
order \mbox{$ \; \sim m \al^7\ $}.

When deriving formulas for corrections to energy levels the ground
state of positronium may be regarded as non-degenerate, because
S-states with \mbox{$ \sigma=0$} and \mbox{$ \sigma=1$}
(where $\sigma$ is the positronium full spin) do not mix.  
For perturbation $V$ depending on full energy of 
a system,  
correction to non-degenerate energy level number $n$ reads
\[   \Delta E_n\ =\ V_{nn}\ \; +\ \;
\sum_{m \neq n}\ \fr{|V_{nm}|^2} {E_n - E_m}\ \; +\
\sum_{k \neq n} \sum_{m \neq n}\
 \fr{V_{nm}\ V_{mk}\ V_{kn}}{(E_m-E_n)\ (E_k-E_n)}\ \; +  \]
\bq\label{eq:DeE}
 +\ V_{nn} \fr{\partial V_{nn}}{\partial E_n}\
+\ \fr{\partial}{\partial E_n}\
\left(\ V_{nn}\ \sum_{m \neq n}\ \fr{|V_{nm}|^2}{E_n - E_m}\
\right)\ +\
 \fr{\partial}{\partial E_n}\
\left(\ \fr{1}{2}\ V_{nn}^2\ \fr{\partial V_{nn}}{\partial E_n}\
\right)\ \; , \ee
 where summation over continuous part of the spectrum, as
well as over discrete one, is implicit;
\mbox{$ E_m,\ E_n,\ E_k\ $} are zeroth-order energy levels; all
matrix elements and their derivatives are taken at
\mbox{$ E = E_n$}.
It may be easily shown that for calculation of \mbox{$\Delta \nu$}
to order $m \al^6$ it is enough to use only three first terms in
(\ref{eq:DeE}). Keeping only terms of the order desired, the recoil
contribution to \mbox{$ \De \nu $} equals (up to the order
\mbox{$m \al^6 $}) 
\bq\label{eq:W}  < \psi |\
V_1 G' V_1 G' V_1 + V_2 G' V_1 + V_1 G' V_2
+ V_3 + V_1 G' V_1 + V_2 + V_1 \;  | \psi > |_{\sigma=0}^{\sigma=1}
  \; \; , \ee
where \mbox{$ | \psi > $} is zeroth approximation to $ |e>\ $, i.e.
solution of equation
\bq\label{eq:psi}
H^{(0)}\ |\psi>\ =\ E_2\ |\psi>\ \; ,  \ee
and $G' = G'(E_2) $ is Green function of equation (\ref{eq:psi})
with the ground state pole subtracted out (the Green function without
the subtraction is denoted as $G$).

As $H^{(0)}$ does not depend on fermion spins,  \mbox{$ | \psi > $}
may be assumed to have the form
\[ | \psi >\ =\ \int\ \fr{d^3 p}{(2 \pi)^3}\
 \fii (\pv)\ | \pv >\ |\chi> \; , \]
where \mbox{$ | \chi > $} is spin part of the wave function. So
to find \mbox{$ | \psi > $} is essentially to find
\mbox{$\fii (\pv)\ $}, i.e. to solve equation
\bq\label{eq:fieq}
(\ T\ +\ V_0\ -\ E_2\ )\ \fii ( \pv )\ =\ 0\ \; \;   \ee
where $T$ is kinetic energy of the two particles (including the 
mass).

Nonrelativistic approximation for (\ref{eq:fieq}) is
 \bq\label{eq:fin}
 (\ T_0\ +\ V_0\ -\ E_0\ )\ \fin ( \pv )\ =\ 0\  ,  \ee
where $T_0$ is nonrelativistic approximation for $T$.
The ground state solution for (\ref{eq:fin}) is
\[ E_0\ =\ 2m\ -\ \fr{\gm^2}{m}\ \;  ,\
\; \; \; \; \; \; \fin(\pv)\ =\ 8\ \gm^{5/2} \pi^{1/2}\
\fr{1}{f_p^2}  \]
(here and below notations
\mbox{$ \gm \equiv\ \al m/2\ ,\ \; f_k \equiv\ \kv^2 +
\gm^2\ \; \; $} are used, for an
arbitrary momentum $\kv$).

Let us write $\fii\ $ as
\[ \fii\ =\ C_0\ \fii_0\ +\ \defi\ \; ,  \]
where \mbox{$ \; \defi\ $} is orthogonal to $\fin$;
then (\ref{eq:fieq}) is rewritten as
 \bq\label{eq:defi}
  (\ T_0 + V_0 - E_0\ )\ \defi\ =\
 (\ -\ \De T\ +\ \De E\ )\ \fii \; \; ,  \ee
where
\mbox{$ \; \De E\ \equiv\ E_2 - E_0\ ,\
\; \De T\ \equiv\ T - T_0\ \; $}. Solution of (\ref{eq:defi})
may be written in the form
 \bq\label{eq:defis}
\defi\ =\ G_0'(E_0)\ (\ \De T\ -\ \De E\ )\ \fii \; \; ,  \ee
where $ G_0' $ is Green function of equation (\ref{eq:fin}) (i.e.
usual nonrelativistic Coulomb Green function) with the ground state pole
subtracted out (corresponding Green function without the subtraction is
denoted as $G_0$).

Relativistic free two-particle
Green function and nonrelativistic free one are
referred to as $S_0$ and $S$:
\[  S_0 ( E')\ =\ ( E' - T_0 )^{-1}\
\; \; \; ,  \]
\[  S (E')\ =\ ( E' - T )^{-1}\ \; \; ;  \]
$G$ and $G_0$ are expressed as expansions
\[ G_0 (E')\ =\ ( E'\ -\ ( T_0 + V_0 ) )^{-1}\ =\
 S_0\ +\ S_0 V_0 S_0\ +\ .....\ \; \;  ,  \]
\bq\label{eq:G} G (E')\ =\ ( E'\ -\ ( T + V_0 ) )^{-1}\ =\
 S\ +\ S V_0 S\ +\ .....\ \; \;  .  \ee
Let us also denote
\[ L\ \equiv\  G_0'\ -\ S_0\ - \; S_0 V_0 S_0\ \; .  \]
It may be shown by iteration of (\ref{eq:defis}) that with
the sufficient accuracy  its solution is
\bq\label{eq:fi}
 \fii\ \approx\ (\ 1\ +\ (S - S_0) V_0\ +\
S V_0 (S - S_0) V_0\ +\ L\ \De T\ )\ |_{E'=E_0}\ \fin\ \; \; . \ee
It is convenient to use in the rest of the paper notations
  \mbox{$  \; < \; \; \; >  \; $} and
  \mbox{$  \; < \; \; \; >_{(n)}  \; , \; $}
 defined so that for any operator $X$
\[ < X > \equiv
  \int \fin^{+} (\pv' ) X(\pv , \pv' , E ) |_{\sigma=0}^{\sigma=1} \fin (\pv)
 \fr{d^3 p}{(2 \pi)^3} \fr{d^3 p'}{(2 \pi)^3}\
\; \; \; , \; \;  \]
and \mbox{$ <X>_{(n)}$} stands for the sum of contributions
\mbox{$ \sim \al^n, \al^n \ln{\al} $} to \mbox{$ <X>$}.

If one rewrites (\ref{eq:W}) using
\[ G'\ \approx\ ( S\ +\ S V_0 S\ +\ L\ )|_{E'=E_0}  \; ,  \]
expression for $\fii$ from $\fin$ (\ref{eq:fi}), and keeping only terms
of the relevant order, one obtains
\bq\label{av}
\De \nu_{rec}\ =\
<\ V_3\ +\ V_2\ +\
U_{V2}\ +\ U_C\ +\ U_M\ +\ U_{MM}\ +\ U_{MCM}\
+\ U_{MMM}\ +\ U_L\ >_{(6)}\ \; ,  \ee
where
\bq\label{av2}
U_{V2}\ =\ (V_2 S V_m + V_2 ( S V_c - S_0 V_0 ))\ +\ h.c.\ \; , \ee
\bq\label{av3} U_C\ =\ V_{1c}\ +\ U_{C2}\ +\ U_{C3}\  \; ,  \ee
\bq\label{av4} U_M\ =\ V_m\ +\ U_{M2}\ +\ U_{M3}\  \; , \ee
\bq\label{av51} U_{MM}\ =\ U_{MM2}\ +\ U_{MM3}\ \; \ee
\bq\label{av52} U_{MCM}\ =\ V_m S V_c S V_m \; , \ee
\bq\label{av6} U_{MMM}\ =\ V_m S V_m S V_m \; \; \; , \ee
\bq\label{av7} U_L\ =\ V_1 L V_1\
+\ (\ V_1 L \De T\ +\ h.c. )\  , \ee
and
\bq\label{av10} U_{C2}\ =\ V_{1c} S V_{1c}\ +\ (\ V_{1c} (S - S_0) V_0\
+ h.c.\ ) \; ,  \ee
\bq\label{av11} U_{M2}\ =\ V_m (S V_c - S_0 V_0)\ +\ h.c.\  ,  \ee
\bq\label{av12} U_{MM2}\ =\ V_m S V_m\ \; , \ee
\bq\label{av13} U_{C3}\ =\ V_0 (S - S_0) V_{1c} (S - S_0) V_0\
 +\ (\ V_{1c} S V_0 (S - S_0) V_0\ +\ h.c.\ )\ +\  \ee
\[ +\ (\ V_{1c} S V_{1c} (S - S_0) V_0\  +\ h.c.\ )\ +\
 V_{1c} S V_0 S V_{1c}\ +\ V_{1c} S V_{1c} S V_{1c}\ \; \; , \]
\bq\label{av14} U_{M3}\ =\ (V_c S - V_0 S_0)\ V_m\ (S V_c -S_0 V_0)\
 +\ (\ V_m S V_c\ (S V_c - S_0 V_0)\ +\ h.c.\ )\ \; \; ,  \ee
\bq\label{av15} U_{MM3}\ =\ (\ V_m S V_m\ (S V_c - S_0 V_0) +\ h.c.\ )\
 \; . \ee

\vspace*{0.5cm}

3. THE METHOD OF THE CALCULATION.

\vspace{0.5cm}

In this section method of calculation of contributions to
\mbox{$ \De \nu_{rec} $},
written out in (\ref{av}) - (\ref{av15}), is presented.
The main idea of this method is just the idea used in other modern works
concerning bound states, though there is some difference in its
implementation. Every contribution to \mbox{$ \De \nu_{rec} $} is divided
essentially into two parts, "soft" and "hard" ones. Here the following is
undermined. Contributions to \mbox{$ \De \nu_{rec} $} are in fact
integrals over loop momenta.
Any contribution under the
consideration is called "hard" if it is determined
(with accuracy required) by region where all the loop momenta
are of order $m$.
Otherwise this contribution is called
"soft". To separate "soft" contributions from "hard" ones, the
integrands are to be
expanded in powers of the momenta. In order to know whether given
term of such expansion contain "soft" contribution, all the momenta should be
set to \mbox{$ \sim \al m$}, and then simplest power counting let
us know whether this momentum area give rise contribution
\mbox{$ \sim m \al^6 $} to  \mbox{$ \De \nu_{rec} $} (for the term under
the consideration). If this is the case, the term considered clearly must
be treated as "soft" contribution. On the other hand it can be proved that in
the present problem absence of the contribution \mbox{$ \sim m \al^6 $} due to
the momenta region mentioned above lead to absence of "soft" contributions of
the relevant order at all, at least for the way of calculation described in
the present paper.

It is clear that there is some freedom in explicit way of the
decomposition into "soft" and "hard" parts, and "soft" part can always be
defined so that its contribution to the integral is easily evaluated
analytically. On the other hand, calculating "hard" part one can put
\mbox{$ E_0=2m $} in the integrands, and the integrals obtained can be
easily evaluated by means of numerical integration.

 Contributions to $\De \nu_{rec}$ are naturally divided into three
sets, different in form and calculation procedure used. These are
"tree" contributions ($<V_{1c}, V_m >_{(6)}$),
"one-loop" (\mbox{$ < V_2, U_{C2}, U_{M2}, U_{MM2} >_{(6)}$}), and
"two-loop" ones
(\mbox{$ < V_3, U_{V2}, U_{C3}, U_{M3}, U_{MM3},
U_{MCM}, U_{MMM}\ >_{(6)}$})
(here words "one-" and "two-loop" imply that the operators inside
 \mbox{$ < \; >_{(6)}$} are integrals over one and two momenta
respectively),
and also term \mbox{$ < U_L >_{(6)}$} that corresponds to graphs having
three loops or more; the last is determined by region of momenta
\mbox{$ \sim m \al $} and can be easily found analytically.

Consider different kinds of the contributions one by one.

Each two-loop contribution has the form
\mbox{$ <X_2>_{(6)}$} where
\bq\label{x2}
X_2 (\pv, \pv', \gm) \; = \;
\int \; \fr{d^3 q_1}{(2 \pi)^3} \; \fr{d^3 q_2}{(2 \pi)^3} \;
\al^3 \; Y_2 (\pv, \pv', \qv_1, \qv_2, \gm)\  . \ee
It is convenient to define the value $n$,  the "divergency power"
of the $X_2$;
namely, \mbox{$ Y_2 \sim \delta^{n-6} $} at 
\mbox{$ p \sim p' \sim q_1 \sim q_2 \sim \gm \sim \delta \ll m\ $}.
All the "two-loop" contributions have  \mbox{$ n \geq 0 \; $};
if \mbox{$ n=0$} then region
\mbox{$ q_1 \sim  q_2 \sim m \al $} contribute to order
\mbox{$ \sim m \al^6$}
and \mbox{$<X_2>_{(6)}$} may contain terms
 \mbox{$\sim m \al^6 \ln{\al}$}. It can be shown that
for all the "two-loop" terms having \mbox{$n >0 $}
the only essential contribution to the integral arises at
\mbox{$ p \sim p' \sim \gm\ ,\
\; \; q_1 \sim q_2 \sim |\qv_1-\qv_2| \sim m\ \; $};
dependence of $Y_2$ on \mbox{$ \pv, \pv' $} in this area can be
neglected that results in
\bq\label{xp}
 <X_2>_{(6)} \approx
 |\varphi_0 (0)|^2 X_2|_{\pv=\pv'=\gm=0}|_{\sigma=0}^{\sigma=1}\
equiv\ <X_2>_p\ , \ee
where
\[  \varphi_0 (0)\ =\
\int \; \fr{d^3 p}{(2 \pi)^3} \; \fin (\pv) \; \; =
\sqrt{\fr{\gm^3}{\pi}}\ . \]
The integral in (\ref{xp}) does not depend on the small parameter
$\al$ and so can be easily computed numerically.
For the sake of brevity notation \mbox{$ < \; >_p $}
defined in (\ref{xp}) is used throughout the paper.

If \mbox{$ n=0 $} the integrand $Y_2$ is to be divided into two
parts: the soft part, denoted as  $Y_{21}$, and the hard one, $Y_{22}$,
\mbox{$ \; \; \; Y_2 = Y_{21}+Y_{22},\ $} so that
 \mbox{$ Y_{21},\ Y_{22}\ $}
have "divergency power" \mbox{$n=0$} and \mbox{$n>0$}
respectively. It may be done in such a way that  $Y_{21}$
equals approximately a uniform function at
\mbox{$ p,p',q_1,q_2, \gm \ll m $}
(the only exception is the contribution of graph  shown at fig.2(b), treated in
the section 4). Similarly
\mbox{$ \; X_2 = X_{21} + X_{22} $} where  \mbox{$ X_{21}, X_{22} $}
 are integrals of form (\ref{x2}) with the $Y_2$ replaced by
  \mbox{$ Y_{21}, Y_{22} $} respectively.  It may be shown that
 with such choice of $Y_{21}$ in all cases met in calculations below
the correct recipe  for the calculation is
\bq\label{x21}
 < X_{2} >_{(6)}\ =\ < X_2-X_{21} >_p\ +\
 < X_{21} >_{(6)} . \ee
The decomposition into $X_{21}$ and $X_{22}$ is evidently non-unique,
and $X_{21}$ may be choosen so as readily to calculate
 \mbox{$ <X_{21}>_{(6)} $} analytically.

In the method described above the $<X_{21}>_{(6)}$ is defined by the integral
which is assumed to converge. Hence the $Y_{21}$ falls at
\mbox{$ q_1, q_2 \gg m $} faster then uniform function having
\mbox{$n=0$}. As the scale of the momenta, at which the original function  $Y_2$
begins to fall rapidly, is \mbox{$ q_1 \sim q_2 \sim m $}, it is
natural to demand the same behavior from the function $Y_{21}$ as well;
in other words, there exist effective "cutoff" of the $Y_{21}$ at
\mbox{$ q_1 \sim q_2 \sim m $}. This "cutoff" reminds regularizations
in nonrelativistic effective theories; however, in the approach described
here there is no need in artificial regulators such as large
regulating mass or dimension different from 4. Similarly to what
explained above, in the rest of the paper all functions looking like
uniform (with $n=0$) at small momenta fall rapidly enough at large ones.

For calculation of "tree" and "one-loop" contributions it is necessary to
find corrections to "leading" terms, which are of first and second
relative order in the $\al$. The method of finding of these corrections is
rather obvious and is shown by the following simple illustration.
Consider calculation of \mbox{$<X>$} where
 \mbox{$X=X(\pv,\pv')$} is sufficiently smooth function that doesn't
depend on $\al$, and expansion of
 \mbox{$ X(\pv,\pv') - X(0,0)$} in \mbox{$ \pv, \pv'$}
starts from third-order terms. Evidently
\[  <X> =\ (A_0\ +\ A_1\ +\ A_2)|_{\sigma=0}^{\sigma=1}\ \; , \]
\[ A_0\ =\ |\varphi_0 (0)|^2\ X(0,0)\ ,  \]
\bq\label{a1}
 A_1\ =\ \varphi_0 (0)\ \int\ \fr{d^3 p}{(2 \pi)^3}\
 \fin (\pv)\ (X(\pv,0) + X(0,\pv) - 2 X(0,0))\ , \ee
\bq\label{x}
 A_2 = \int
\fr{d^3 p}{(2 \pi)^3} \fr{d^3 p'}{(2 \pi)^3}
\fin^{+} (\pv') \; (X(\pv , \pv') - X(\pv,0) - X(0, \pv') + X(0,0))
\; \fin (\pv)\  . \ee
It is easily seen that when calculating three first terms of
expansion of \mbox{$ <X> $} in $\al$,
in (\ref{a1},\ref{x})
\mbox{$ \fin(\pv) $} may be replaced by
\mbox{$ \; 8 \gm^{5/2} \pi^{1/2}\ p^{-4}\ $};
hence \mbox{$ \; A_{0,1,2} \; $} exactly have orders
\mbox{$ \al^3 \; , \; \al^4 \; , \; \al^5 \; \; $} respectively.

Each "one-loop" contribution has the form \mbox{$ <X_1>_{(6)} $} where
\bq\label{x1}
 X_{1} (\pv, \pv', \gm) \; = \;
\int \; \fr{d^3 q}{(2 \pi)^3} \;
\al^2 \; Y_{1} (\pv, \pv', \qv, \gm)\ ; \ee
Let $n$ to be defined so that
\mbox{$ \; Y_1 \sim \delta^{n-2} \; $}
at \mbox{$ p \sim p' \sim q \sim \gm \sim \delta \ll m\ $}
(if one writes
\mbox{$X_1 S_0 V_0 $} as the loop integral, the $n$ is its
"divergency power";
\mbox{$ n=0 $} imply that the region \mbox{$ q \sim m \al $}
contributes to order \mbox{$ \sim m \al^6 $},
and may cause terms \mbox{$ \sim m \al^6 \ln{\al}$}) to exist.

Let us write \mbox{$Y_1$} for any given contribution as sum of "soft"
and "hard"
parts, \mbox{$ Y_1=Y_{11} + Y_{12}\ $}, where
$Y_{11}$ and $Y_{12}$ has \mbox{$n \leq 0$} and \mbox{$n>0$}
respectively (similarly \mbox{$ X_1=X_{11}+X_{12}$}).
It may be done so that $Y_{11}$ at \mbox{$ p,p',q, \gm \ll m $}
equals approximately a uniform function (the only exception is the
contribution of graph  shown at fig.2(a), which is treated in the section 4).
It can be proved that under such choice of $Y_{11}$ contributions coming
from $Y_{12}$ can be obtained, for all terms to treat, using formula
like (\ref{a1}), which results in
\bq\label{x11}
 <X_1>_{(6)}\ =\  <
((X_1 - X_{11})\ -\ (X_1 - X_{11})_0 )\ S_0 V_0\
 \; +\ h.c.\ >_p\ \;
 +\ < X_{11} >_{(6)}\ \; ;   \ee
in (\ref{x11}) the notation $( \; \; \; )_0 \; , \; $ is used, defined
so that
\mbox{$ (X)_0 (\pv,\pv') \equiv X(\pv,\pv')|_{\pv=\pv'=0}\ $}.
Evidently $Y_{11}$ can always be choosen so that second term in
(\ref{x11}) is easily evaluated analytically.

"Tree" contributions to $\De \nu_{rec}$ are calculated in the way
just like one described above. There are only two "tree" contributions
to be found: \mbox{$ <V_m>_{(6)}$} and \mbox{$ <V_{1c}>_{(6)}$}.
Each of them is divided into "soft" and "hard" parts (the explicit way of
this partition is described in the section 4), and then the "hard"
contributions are evaluated according to (\ref{x}).

In the section 4 separation of "soft" contributions from
$\De \nu_{rec}$ is done
using formulas (\ref{x21}),(\ref{x11}),(\ref{x}), that results in
\bq\label{V3av}
 <V_3>_{(6)}\ =\ <V_3 - W_{V3} >_p\ +\ E^S_{V3}\  \; ,  \ee
\bq\label{V2av}
 <V_2+U_{V2}>_{(6)}\ =\ < V_2 S V_m + h.c. >_p\ +\
 < (V_2 S V_c - W_{V2} ) + h.c. >_p\ +\ E^S_{V2}\ , \ee
\bq\label{U0av}
 <U_C>_{(6)}\ =\
< V_c S V_c S V_c - W_C >_p\ +\ E_C^S\ ,  \ee
\bq\label{U1av}
<U_M>_{(6)}\ =\
< V_c S V_m S V_c - W_{M1} >_p \; + \;
< ( V_m S V_c S V_c - W_{M2} ) + h.c. >_p \; + \; E^S_M\ \;
 ,  \ee
\bq\label{UMMav}
<U_{MM}>_{(6)}\ =\
< ( V_m S V_m S V_c - W_{MM} ) + h.c. >_p\ +\ E^S_{MM} \
 \; ,  \ee
\bq\label{UMCMav}
<U_{MCM}>_{(6)}\ =\
< V_m S V_c S V_m - W_{MCM} >_p\ +\ E^S_{MCM} \
 \; ,  \ee
\bq\label{U3av}
<U_{MMM}>_{(6)}\ =\ < V_m S V_m S V_m >_p\
 ,  \ee
here $ W_{V3},\ W_{V2},\ W_C,\
W_{M1},\ W_{M2},\ W_{MM},\ W_{MCM}\ $ are operators associated with
the
soft contributions, which have relatively simple form,
\mbox{$E^S_{V3},\ E^S_{V2},\ E^S_M,\ E^S_{MM},\ E^S_{MCM},\ E^S_C $}
are analytically found values of "soft" contributions to
\mbox{$\De \nu_{rec}$}; these values and operators are written
explicitly in the section 4.

Remind that operators
\mbox{$ \; V_0+V_1 \; , \; V_2 \; , \; V_3 \; $}
are given by sums of all the tree, one-loop, and two-loop irreducible
noncovariant graphs respectively. Hence first terms in
\mbox{$ \; <\ \; >_p\ $} in (\ref{V3av}) - (\ref{U3av})
are also given by sums of some noncovariant graphs (reducible ones, except
$V_3$). The sum of them is equal to the sum of expressions associated with all
two-loop graphs of the recoil type. The operator corresponding to this sum is
referred to as $W$. For convenience of references, the following
notations are also used: sum of all second terms in
 \mbox{$  <\ \;  >_p \; $}
in (\ref{V3av}) - (\ref{U3av}) is denoted as $W_0$, and sum
of all the soft contributions as $E^S$, hence
\[ \De \nu_{rec}\ =\ <\ W - W_0 >_p\ +\ E^S\  \; ;  \]
also, notations \mbox{$ W',\ W_0'\ $} are used, defined according to
\[  W, W_0(\pv, \pv', \gm )|_{\pv=\pv'=\gm=0}\
=\ \int\ \fr{d^3 p_1}{(2 \pi)^3}\ \fr{d^3 p_2}{(2 \pi)^3}\
W',\ W_0'\ (\pv_1, \pv_2 ) \; .  \]

A sum of contributions to $W'$, arising from all noncovariant graphs with the
same topological structure (taking into account the difference of
magnetic and Coulomb quanta), is equal to integrand associated with
usual covariant Feynman graph (after the integration over zero components
of the loop momenta).  Hence the $W'$ can be determined as sum of
contributions of
all {\it covariant} two-loop graphs of the recoil type, and there is no need
to know contributions to the $W'$ arising from separate noncovariant graphs.

In the above the gauge used for noncovariant and covariant graphs was
undermined
to be the same. However sum of contributions of all covariant graphs
to the $W'$ is
independent of gauge, and so in actual calculations any gauge may be
choosen. This is quite natural; in terms of the NRQED (or the effective
nonrelativistic Hamiltonian approach) \mbox{$ <W-W_0>_p$} is equal,
besides some overall factor, to contribution \mbox{$ \sim \al^3 $}
to the constant of pointlike fermion-fermion interaction (if
an appropriate regularization is used), which is determined from
scattering amplitudes on the mass shell.

\vspace*{0.5cm}

4. THE "SOFT" CONTRIBUTIONS.

\vspace*{0.5cm}

In this section the general method of separation of "soft"
contributions, described  
above, is applied to various terms of  
(\ref{av}) - (\ref{av15}) and to contributions of various graphs.  

There are some preliminary remarks concerning the calculation technique
employed.
First, every term contributing to $W$ and $W_0$ evidently may be
replaced by its average over positronium polarizations (or, in other
words, over directions of the full spin).
This simplifies the calculations considerably. For contributions to $W$ this
average is found in program way, as $1/3$ of sum over polarizations.
For contributions to $W_0$ the averaging is performed using the following
formulas (to the r.h.s. of which terms independent of the full spin may
be added):
\bq\label{nps2}
 ( \so \aav )\ ( \sd \bv )\ \; \arr\ \;
 \fr{1}{3}\  (\so \sd\ )\ (\aav \bv) \; , \ee
\bq\label{nps1}
 (  \so \aav )  \; ( \sd \bv ) \;
\left( \; \so \sd\ -\ \fr{1}{\qv^2}\ ( \so \qv )\
( \sd \qv)\ \right) \; \arr\ \;
-\ \fr{1}{3}\ \so \sd
\; ( \aav \bv + (\aav \qv ) (\bv \qv ) / \qv^2 \; )  \; , \ee
\bq\label{nps3}
\left( \; \so \sd\ -\ \fr{( \so \qv )\ ( \sd \qv)}{\qv^2}\
\right) \;
\left( \; \so \sd\ -\ \fr{( \so \kv )\ ( \sd \kv)}{\kv^2}\
\right) \;
\; \arr\ \; \so \sd\
\left( \; -1\ +\ \fr{1}{3}\ \fr{(\qv \kv)^2}{\qv^2 \kv^2}\
\right) \; . \ee
Second, let me describe calculation of contributions associated with
retardation. These contributions have "ultrasoft" parts,  
for which virtual photon momenta  
\mbox{$ \sim m \al^2$} are essential. These contributions are contained in
\mbox{$ \; <V_m>_{(6)},  <V_2>_{(6)}, <V_3>_{(6)} \; $}.
Considering them, it is convenient to replace these operators by
their spin-spin parts averaged over positronium polarizations, as
described above. First nonrelativistic approximations (in the sense
of the power expansion in momenta) for averaged contributions of
graphs 2(a) and 2(b)
to $V_2$ and  $V_3$ are
$V_2^{ret}$ and $V_3^{ret}$,
\bq\label{V2ret}
  V_2^{ret} (\pv, \kv )\ =\
\int\  \fr{d^3 q}{(2 \pi)^3}\  \al^2\
\fr{4}{3} \fr{\pi^2}{m^2}\ \sisi \fr{q}{q'^{ \; 2}}\ \;
Y_1^{ret} (\pv, \kv, \qv )\
\; , \ee
\bq\label{V3ret}
 V_3^{ret} (\pv, \kv )\ =\
\int\ \fr{d^3 q_1}{(2 \pi)^3}\ \fr{d^3 q_2}{(2 \pi)^3}\
\al^3\
\fr{16}{3}\ \fr{\pi^3}{m^2}\ \sisi\ \fr{q}{q_1^2 q_2^2}\ 
Y_2^{ret} (\pv, \kv, \qv_1, \qv_2 )\
\; , \ee
where
\bq\label{Y1ret}
Y_1^{ret} (\pv, \kv, \qv )\ =\
\fr{1}{ q + (f_p + f_{k_1})/2}\ \;
\fr{1}{ q + (f_k + f_{p_1})/2}\ \;
 \;  ,  \ee
\bq\label{Y2ret} Y_2^{ret} (\pv, \kv, \qv_1, \qv_2 )\ =\
\fr{1}{(q+(f_p+f_{k_2})/2)}\
\fr{1}{(q+(f_{p_1}+f_{k_1})/2)}\
\fr{1}{(q+(f_{p_2}+f_k)/2)}\
\ee
(notations for the momenta are shown at fig. 2); just the same contributions
arise due to graphs obtained from graphs 2(a) and 2(b) by the
time reversal. 

Spin-spin part of $V_m$, averaged over positronium polarizations,
is denoted below as $V_s$. First nonrelativistic approximation to
\mbox{$ V_s (\pv,\kv) $} is
\bq\label{Vmret}
 V_m^{ret} (\pv, \kv)\ =\ V_{m0}\ \fr{q}{q+(f_p+f_k)/2}\ \; ,
\ee
where
\[ V_{m0}\ =\ V_s|_{\pv=\kv=\gm=0}\ =\ \fr{2}{3} \fr{\al \pi}{m^2} \sisi\
\; , \; \; \; \;
 \qv \equiv \kv - \pv \; \; . \]

It is convenient to consider
\mbox{$ \; <V_m^{ret}>_{(6)},  <V_2^{ret}>_{(6)},
<V_3^{ret}>_{(6)} \; $} together and to calculate their overall
contribution to
 \mbox{$ \De \nu_{rec}$}; this way of the calculation was employed,
in particular, in \cite{Pa} and \cite{Elh}. The way of the
calculation, used in the present paper is, in some sense, even
shorter. Each of operators
\mbox{$ \; V_m^{ret}, V_2^{ret}, V_3^{ret}\  $} is divided into
two parts:
\[  V_m^{ret} (\pv, \kv)\ =\ V_{m0}^{ret} (\pv, \kv)\ +\
 V_{m0}' (\pv, \kv)\ \; \; ,  \]
\[  V_2^{ret} (\pv, \kv)\ =\ V_{20}^{ret} (\pv, \kv)\ +\
 V_{20}' (\pv, \kv)\ \; \; ,  \]
\[  V_3^{ret} (\pv, \kv)\ =\ V_{30}^{ret} (\pv, \kv)\ +\
 V_{30}' (\pv, \kv)\ \; \; ,  \]
where
\bq\label{Vn0}
 V_{m0}^{ret}(\pv, \kv)\ =\ \fr{1}{2} V_{m0}\
\left[ \; \fr{q}{q\ +\ f_p}\ +\ \fr{q}{q\ +\ f_k} \; \right]\ \; , \ee
and $V_{20}^{ret}$, $V_{30}^{ret}$ are obtained from 
$V_{2}^{ret}$, $V_{3}^{ret}$ by replacing 
\mbox{$ Y_1^{ret}, Y_2^{ret} \arr Y_{10}^{ret}, Y_{20}^{ret} $}
where  
\bq\label{Y10ret} Y_{10}^{ret} (\pv, \kv, \qv)\ =\
  \fr{1}{2}\   \left(\  \;
\fr{1}{ ( q + f_k )\ ( q + f_{k_1} )}\ \; +\ \;
\fr{1}{ ( q + f_p )\ ( q + f_{p_1}  )}\  \; \right)\
\; \; \; , \ee
\bq\label{Y20ret} Y_{20}^{ret} (\pv, \kv, \qv_1, \qv_2 )\ =\
\fr{1}{2}\  \left( \;
\fr{1}{(q+f_p)(q+f_{p_1})(q+f_{p_2})} \; + \;
\fr{1}{(q+f_{k})(q+f_{k_1})(q+f_{k_2})} \; \right)\ \;
\; .  \ee
In the region where all the momenta are \mbox{$\sim m \al$} 
\[  V_{m}^{ret}\ \approx\ V_{m0}^{ret}\ \; ,\ \; \; \; \;
Y_1^{ret}\ \approx\ Y_{10}^{ret}\ \; ,\ \; \; \; \;
Y_2^{ret}\ \approx\ Y_{20}^{ret}\ \; .  \]
If one writes \mbox{$ \; V_{20}', V_{30}' $} as integrals like
(\ref{V2ret},\ref{V3ret}),
the integrands
in them have powers, being expanded in the momenta, greater by 2 than
those of \mbox{$ Y_1^{ret}\ ,\ Y_2^{ret}\ $}; also \mbox{$ \; V_{m0}'$} has
power, being expanded in the momenta, greater by 2 than the power of
$V_m^{ret}$.
Due to this reason
\mbox{$ \; <V_{20}'>_{(6)},\ <V_{30}'>_{(6)}\  $}
consist of "hard" contributions only.
As for 
\mbox{$ \; <V_{m0}'>_{(6)},\  $} it does contain soft contributions. However
it may be easily calculated in just the way other soft contributions to
\mbox{$ <V_{m}>_{(6)}\  $} are calculated,
and is actually evaluated combined with them in sec.4.3.

Now turn to contributions of
\mbox{$ V_{m0}^{ret}, V_{20}^{ret}, V_{30}^{ret} $}.
It is easy to see that their full contribution is equal to 0:
\bq\label{ret}
 <V_{m0}^{ret}>_{(6)}\ +\  2 <V_{20}^{ret}>_{(6)}\ +\
2 <V_{30}^{ret}>_{(6)}\ =\ 0\
\; \; . \ee
This may be shown by a short direct calculation; however (\ref{ret})
is evident immediately due to the following reason. Consider
calculation of the recoil
contribution \mbox{$ \sim \al^6 m^2/M $} to ground state
hyperfine splitting of the "hydrogen", i.e. particle of mass
\mbox{$ m/2 $} moving around particle of large mass $M$.
In this calculation
\mbox{$ V_m^{ret} (\pv, \kv )\ $}, for instance, must be replaced by
\mbox{$ 2 m/M\ V_{m0}^{ret} (\pv, \kv )\ $}; entirely
\mbox{$ < V_m^{ret}+ 2 V_2^{ret}+2 V_3^{ret} >_{(6)} $}
is replaced by \newline 
\mbox{$2 m/M\ <V_{m0}^{ret}+ 2 V_{20}^{ret}+2 V_{30}^{ret}>_{(6)}$},
as may be easily seen.
On the other hand, it is well known that when calculating
hyperfine splitting in first  
order in $m/M$ the magnetic interaction may be regarded as instant,
which immediately leads to
\bq\label{ret0}
 <\ V_{m0}^{ret}\ +\ 2 V_{20}^{ret}\ +\ 2 V_{30}^{ret}\ >_{(6)}\
=\ <\ V_{m0}\ >_{(6)}\  =\ 0\
\; \; . \ee

The last point to discuss is calculation of \mbox{$ <U_L>_{(6)}$}, that is
essentially the contribution of graphs containing Coulomb ladder with
three or more
loops. This contribution to \mbox{$ \De \nu_{rec} $}
arises completely, with the accuracy required, from the region where all
the momenta in the loops are of order \mbox{$ m \al $}, and can be
easily evaluated analytically. There are two kinds of such contributions,
namely, graphs having one or two magnetic quanta. The contribution of graphs
having two magnetic quanta is
\[ E^S_{LMM}\ \equiv\ <\ V_m L\ V_m >_{(6)}\ \;   \]
and does not depend, in the order required, on details of
formalism used; for the first time it has been found in \cite{CL78}.
Method of its evaluation, used in the present paper, coincides mainly
with the method used in \cite{BYG} (and involves evaluation in
coordinate representation); the result obtained agrees with those
found earlier and equals 
\bq\label{RMM}
E^S_{LMM}\ =\ \left(\ \fr{791}{864}\ -\ \fr{\pi^2}{18}\ \right)\ m \al^6\
\approx\ 0.3672\ m \al^6\  \; .  \ee

The contribution of graphs having one magnetic photon does depend on details
of formalism used. The value calculated in the present paper is
\[ E^S_{LM}\ \equiv\ <\ (\ V_m L\ (\ V_{1c}+\De T\ ))\ +\ h.c. >_{(6)}\
 \; .  \]
This value was evaluated in two ways: in coordinate space (by method
close to that of \cite{BYG}), and by immediate integration over
the momenta, using explicit form of $L(\pv,\kv)$, quoted, for
instance, in \cite{CL78}:
\bq\label{L}  L(\pv,\kv)\ =\
\fr{- 4 \pi m^4 \al^3}{f_p^2 f_k^2}\ \left(\
\fr{5}{2}\ -\ 4 \fr{\gm^2}{f_p}\ -\ 4 \fr{\gm^2}{f_k}\
+\ \fr{1}{2} \ln{A}\
+\ \fr{2A - 1}{\sqrt{4A-1}}\ \arctan{\sqrt{4A-1}}\ \right)\
 \; \; ,  \ee
where
\[ A\ =\ \fr{f_p f_k}{4 \gm^2 q^2} \; \; .  \]
The results obtained by both methods are the same and equal to
\bq\label{RM}
E^S_{LM}\ =\ \fr{1}{64}\ m \al^6\
\approx\ 0.01562\ m \al^6\  \; .  \ee
Finally, addition (\ref{RMM}) and (\ref{RM}) together results in
\[ <\ U_L\ >_{(6)}\ \approx\ 0.3728 m \al^6\  \; .  \]


\newpage
4.1. Irreducible Graphs.

For calculation of \mbox{$<V_3>_{(6)}\ $} it is necessary first to
separate from $V_3$ the term $V_{30}$ which is an integral having
the "divergency power" $n=0$, and to use formula (\ref{x21}) where
$X_{21}$ should be set equal to
$V_{30}\ $; the result has the form (\ref{V3av}) with
 \[  W_{V3}\ =\ V_{30}\ ,\  \;
E^S_{V3}\ =\ <V_{30}>_{(6)}\ . \; \]
There exist only three essentially different graphs contributing to
$V_{30}$: these are graphs depicted at fig. 3(a),3(b) and 2(b). However
contribution to $V_{30}$, due to sum of graphs 3(a) and 3(b), may be set equal to
zero. The matter is that this sum
 in the region of small loop momenta reduces
effectively to the graph  3(c), in which effective two-photon "seagull"
vertex does not depend on the spins. This vertex appears as sum of
subgraphs 3(d) and 3(e) and corresponds to term
\mbox{$ \al {\bf A}^2 /(2m) $} (where ${\bf A}$ is the vector potential)
in the nonrelativistic Hamiltonian of particle in magnetic field.
Thus $V_{30}$ arises from the only graph  2(b)
and may be choosen to be \mbox{$ 2 V_{30}^{ret}\ $} which results in
\mbox{$ E^S_{V3} = 2 <V_{30}^{ret}>_{(6)}\ $}; in fact there is no
need to evaluate this value, as mentioned above.

Similarly \mbox{$ \; <V_2>_{(6)}\ $} is calculated using
(\ref{x11}) with \mbox{$X_{11}=V_{20}$} where
 $V_{20}$ is part of $V_2$, having \mbox{$ n=0\ $}.
Noting that
\mbox{$ <U_{V2}>_{(6)}=<U_{V2}>_p\ ,\ $}
as is easily seen, one finds (\ref{V2av}) where
\[ W_{V2}\ =\ (\ V_{20}\ +\ ( V_2 - V_{20})_0\ )\ S_0 V_0\
 \; \; \; ,\  \; \; \; \;
 E^S_{V2}\ =\  < V_{20} >_{(6)} \; \; \; . \]
 Consider contributions to $V_{20}$. Graphs having two Coulomb quanta
do not contribute to $V_{20}$.    
Among graphs having two magnetic quanta there are those with
\mbox{$n=-1$} (graphs depicted in fig. 3(f),(g)), and several graphs with
\mbox{$n=0$}. However their contributions to $V_{20}$ cancel each other, just
as described above for $V_{30}$; for instance, sum of graphs shown in
fig. 3(f) and 3(g) reduces effectively in the region of small momenta
to the graph
3(h), in which effective seagull vertex does not depend on spins.
Graphs having one magnetic and one Coulomb quanta, contributing to
$V_2$, are divided into 4 sets belonging each to one of covariant graphs
(A),(B), shown in fig.4, and the graphs obtained by the time reversal. 
It is convenient to consider graphs
including electron-positron pairs and those without pairs separately.

 Contributions of graphs including pairs to $V_{20}$ may be choosen to be
\bq\label{V20pair}
  V_{20}^A (\pv, \pv')\ =\ V_{20}^B\ =\ .... =\
\int\ \fr{d^3 q}{(2 \pi)^3}\
\fr{2 \al^2 \pi^2}{3 m^3} \sisi
  \fr{1}{q^2 q'^2}\
((q^2 + q'^2)\ R_q\ -\ k^2 R_k\ )\ \; \; ,  \ee
(the momenta notations are shown in fig. 4,
\mbox{$ \; \; \kv \equiv \pv' - \pv \; $},
and \mbox{$ \; \; R_k \equiv m^2/(m^2+k^2)\ $} for an arbitrary $\kv$ ).
It is easy to find their overall contribution to $E^S_{V2}$, that equals 
\[ E^S_{P}\ =\ \fr{2}{3}\ m \al^6\ \ln{\al}\  .  \]
The only graphs without pairs, contributing to $V_2$, are the graph shown in
fig. 2(a), discussed above, and the similar graph
obtained from 2(a) by the time reversal. 
Their contribution to $V_{20}$ is choosen to be
$ 2 V_{20}^{ret}$. So
\bq\label{T1D}
 E^S_{V2}\ =\ E_P^S\ +\ 2 <V_{20}^{ret}>_{(6)}\ \; .   \ee
The second term in (\ref{T1D})
need not in fact be evaluated, as explained above.

\vspace{1.5cm}

4.2. Graphs with one magnetic photon.

\vspace{0.5cm}

Consider calculation of \mbox{$<U_M>_{(6)}\ ,\ $} i.e. contribution to
\mbox{$ \De \nu_{rec}$} due to
reducible graphs with one magnetic photon, 
except corresponding contribution to \mbox{$ <U_L>_{(6)}\ $},
 considered above.

As discussed above, it is convenient to calculate
\mbox{$<V_s>_{(6)}$} instead of \mbox{$<V_m>_{(6)}$}, where
 $V_s$ is spin-spin part of $V_m$,
averaged over positronium polarizations. Let us write $V_s$ as an
"expansion in powers of momenta":
\bq\label{Vs}
 V_s\ =\ V_{m0}^{ret}\ +\ V_{m1}\ +\ V_{m2}\ \; , \ee
where $V_{m0}^{ret}$ is defined in (\ref{Vn0}),
\mbox{$ \; V_{m1} (\pv, \kv, \gm )\ \approx\ V_s -V_{m0}^{ret}\ $} at
\mbox{$ \; p \sim k \sim \gm \ll m\ $}, and
\mbox{$V_{m1} $} at \mbox{$ \; p,k, \gm \ll m\ $}
equals approximately a uniform function of second power.
It can be shown that
\mbox{$ \; <V_{m2}>_{(6)}\ $} can be evaluated using (\ref{x}) with
\mbox{$ X\ =\ V_{m2}\ $}.
Actually there is no need to evaluate
 \mbox{$ <V_{m0}^{ret}>_{(6)}\ $},  as mentioned above.
The value \mbox{$ <V_{m1}>_{(6)}\ $} can be easily found in analytic
form, if one uses appropriate choice for \mbox{$ \; V_{m1} \; $}.
It is convenient to choose
\bq\label{Vm1} V_{m1}(\pv, \kv)\ =\ \fr{1}{2 m^2}\ V_{m0}\
 \left[\ \fr{p^2\ k^2}{q^2 }\ (\ R_p\ +\ R_k\ )\
 -\ \fr{1}{q^2 }\ ( p^4 R_p\ +\ k^4 R_k )\
 -\ ( p^2 R_p\ +\ k^2 R_k )\ \right]\ \; ,  \ee
where \mbox{$(\qv \equiv \pv - \kv )\ . \; $}

In the lowest order in momenta \mbox{$ \; V_m \approx\ V_{mb}\ $}
where $V_{mb}$ is part of the Breit Hamiltonian due to magnetic
photon exchange.  
The value \mbox{$ <U_{M2}>_{(6)}$} is calculated using
(\ref{x11}) with
\newline
\mbox{$ X_{11} = V_{mb} (S_0 V_{11}' + S_{11} V_0) + h.c.
\; , \; $} where
\mbox{$ \; S_{11}, \; V_{11}' \; \; $} are "first terms of expansion"
of \mbox{$ S - S_0\ $} and \mbox{$ V_c - V_0\ $} in the momenta
(namely,
\mbox{$ V_{11}'( \pv, \kv)\ \approx V_c - V_0 ,\
S_{11}(\pv) \approx S-S_0$}
\newline
at \mbox{$ \; p \sim k \sim \gm \ll m\ $},
 and \mbox{$ S_{11}\ ,\ V_{11}'\ $}
 at \mbox{$ \; p,k \ll m\ $} equal approximately to uniform functions
of zero power).

The value \mbox{$ <U_{M3}>_{(6)} \; $} is calculated using
(\ref{x21}) with
\newline
\mbox{$ \; X_{21} = V_{mb} S_0 V_0 (S_0 V_{12}' + S_{12} V_0)
+ h.c. \; $}, where
\mbox{$ \; S_{12} (\pv), V_{12}'(\pv, \kv)\ \approx
S_{11}(\pv), V_{11}'(\pv, \kv)\ \; $}
\newline
at \mbox{$ \; p,k \ll m \; $}
(but generally  \mbox{$ \; S_{12}, V_{12}'  \neq
S_{11}, V_{11}' \; ; \; \; $}
the explicit way of the decomposition of
\mbox{$ \; S - S_0 \; $} and \mbox{$ \; V_c - V_0 \; $}
is determined by the purpose of convenience of subsequent calculations).

It is easy to see that in \mbox{$ \; X_{11},\ X_{21}\ $} defined in
such manner it
is enough to take into account only spin-independent parts of
\mbox{$V_{11}',V_{12}'$}
(denoted as \mbox{$V_{11},V_{12}$} respectively) and
spin-spin part of $V_{mb}$, that can be replaced by its average over
positronium polarizations, i.e. $V_{m0}$.
So \mbox{$ \; <U_{M2}>_{(6)}\ $} and
 \mbox{$ \; <U_{M3}>_{(6)} \; $}
can be evaluated using
\newline
\mbox{$ X_{11} = V_{m0} (S_0 V_{11} + S_{11} V_0) + h.c.\  , \; $}
\mbox{$ X_{21} = V_{m0} S_0 V_0 (S_0 V_{12} + S_{12} V_0) + h.c.\ .\ $}

It is convenient to choose
\newline
\mbox{$ \; V_{12} (\pv,\kv) = \al \pi/m^2 R_p =
\al \pi/m^2\ (1+p^2/m^2)^{-1} \; \; $};
\newline \mbox{$ \; V_{11} \; $}
is defined so that
\mbox{$ \; V_{11} (\pv,\kv) = V_{11} (\pv,0) \; ; \; $}
under this condition an explicit form of \mbox{$ \; V_{11}\ $}
 does not enter final expressions.

It is easy to find that $S_{11}, S_{12}$ may be set to
\bq\label{s}
 \; S_{1i} (p)\ =\ S_{1i} (p)|_{\gm=0}\
+\ \fr{1}{2}\ \fr{\gm^2}{f_p}\ -\ \fr{1}{4}\ \fr{\gm^4}{f_p^2}
 \; \; , \; \; \; \; \; \; (\ i=1,2\ )\ \; ; \ee
first term in (\ref{s}) is choosen to be
\mbox{$ ( S-S_{0} )|_{( \gm=0 )} $} for $S_{11}$ and
\mbox{$-R_p/4$} for  $S_{12}$.

After a simple calculation
\mbox{$<U_M>_{(6)}$} is found to have the form (\ref{U1av}) with
\[ E^S_M\ =\ m \al^6\
\left(\ - \fr{1}{3} \ln{\al}\ +\ \fr{53}{192}\ \right)\
+\ <V_{m0}^{ret}>_{(6)}\
\; , \]

\[ W_{M1}\ =\ ((V_c S V_m)_0 S_0 V_0 + h.c.)\ +\
V_0 S_0 (V_{m0}^{ret} + V_{m1}) S_0 V_0\
-\ ((V_0 S_0 (V_{m0}^{ret} + V_{m1}))_0\ S_0 V_0 + h.c.)\  , \]
\[ W_{M2}\ =\ V_{m0} S_0 V_0 (S_0 V_{12} + S_{12} V_0)\
+\ (\ V_m S V_0 + (V_m S (V_c-V_0) )_0\ )\ S_0 V_0\   \]
(when deriving $W_{M2}$ the term
\[ (\ (V_m-V_{m0}) S_{11} V_0\ -\ ((V_m-V_{m0}) S_{11} V_0)_0\
)\ S_0 V_0\ \; \; ,  \]
was added to it, which causes no additional contribution of
order $m \al^6$).

\vspace{1.0cm}

4.3. Coulomb graphs.
\vspace{0.5cm}

The Coulomb contribution due to reducible graphs  
 to \mbox{$ \De \nu_{rec} $} is equal to \mbox{$ <U_C>_{(6)} $}. 
This value was found in the style described above:
\mbox{$ <V_{1c}>_{(6)},\ <U_{C2}>_{(6)},\ <U_{C3}>_{(6)}$}
(i.e. separate terms of \mbox{$ <U_C>_{(6)} $})
were evaluated using (\ref{x},\ref{x11},\ref{x21})
respectively.
The calculation is simplified by the fact that in lowest (i.e.
zeroth) order in momenta $V_{1c}$
contains only spin-independent terms, and besides that
\[  V_{1c}^S ( \pv , \kv ) |_{\pv=0}\ =\
V_{1c}^S ( \pv , \kv ) |_{\kv=0}\ =\ 0
\; \; \; ,   \]
where $V_{1c}^S$ is spin-spin part of $V_{1c}$.

It is easy to see that \mbox{$ <U_C>_{(6)}$} is determined by
the formula (\ref{U0av}), in which
\bq\label{cv}
 W_C = V_0 S_0 V_{1c}' S_0 V_0\ \; ,\
\; \; \; \; \; \; E_C^S\ =\ < V_{1c}' >_{(6)}
\; \; \; ;  \ee
here \mbox{$ V_{1c}' $} is first nonvanishing term of
"expansion" of \mbox{$ V_{1c}^S $} in momenta
(exactly,
\mbox{$ V_{1c}'( \pv, \kv) \approx V_{1c}^S(\pv, \kv)$}
at \mbox{$ p \sim k \ll m $},
and \mbox{$ V_{1c}'$}  at \mbox{$ \; p,k \ll m\ $}
equals approximately a uniform function of second power).

In (\ref{cv})  one can replace \mbox{$ \; V_{1c}' $}
by its average over positronium polarizations (denoted as
\mbox{$ \; V_{1c}'' $}); it is convenient to set
\[ V_{1c}'' ( \pv , \kv )\
=\ \fr{1}{24}\ \so \sd\ \; \fr{\al \pi}{m^2}\ \left(\
-\ \fr{1}{2} \left( \fr{p^4}{q^2} - p^2 - \pv \kv \right)\ R_p\
+\ ( \pv \leftrightarrow \kv )\
+\ \fr{p^2 k^2}{q^2}\ R_p \; \right)
\; \; \; ; \]
using it, one easily obtains
\[ E_C^S =
 -\ \fr{1}{48}\ m \al^6\ \left(\ \ln {\al}\ +\ \fr{1}{4}\ \right)\
  \; . \]

\vspace{0.5cm}

4.4. Graphs with two magnetic quanta.

\vspace{0.5cm}

Turn to calculation of
\mbox{$ \; <U_{MM2} + U_{MM3} + U_{MCM}>_{(6)}\ $}, i.e.
contribution of reducible graphs with two magnetic quanta,
except corresponding contribution to \mbox{$ <U_L>_{(6)}\ $},
 considered above.
The value
\mbox{$ <U_{MM2}>_{(6)}\ $} is found using (\ref{x11}) with
 \mbox{$ X_{11}\ = V_{mb}\ S_0\ R\ V_{mb}\ $}, where 
\mbox{$ \; R(\pv, \pv')\ =\ (2 \pi)^3 \delta^3 (\pv - \pv')\ R_p $}.  
Similarly, the value \mbox{$ <U_{MCM}>_{(6)}\ $}
is found using (\ref{x21}) with
\mbox{$ \; X_{21}\ = V_{mb}\ S_0 V_0 S_0\ R\ V_{mb}\ . $}
It is easy to see that
\mbox{$ <U_{MM3}>_{(6)}\ =\ <U_{MM3}>_p\ $}.
The results have form (\ref{UMMav}),(\ref{UMCMav}) with
\[ W_{MM}\ =\ V_{mb}\ S_0\ R\ V_{mb}\ S_0 V_0\ +\
(\ V_m S V_m -  V_{mb} S_0 R V_{mb}\ )_0 S_0 V_0\  , \]
\[ W_{MCM}\ =\ V_{mb}\ S_0 V_0 S_0\ R\ V_{mb}\ ,  \]
\[ E^S_{MM}\ =\ < V_{mb} S_0 R V_{mb} >_{(6)}\ \; ,\ \; \; \; \; \;
\; \;
 E^S_{MCM}\ =\ < V_{mb} S_0 V_0 S_0 R V_{mb} >_{(6)}\ . \]
Spin-orbital part of $V_{mb}$ does not contribute to
 \mbox{$E^S_{MM}$} and \mbox{$E^S_{MCM}$}.
Spin-spin and spin-independent contributions to
$V_{mb}$ are denoted as $V_{sb}$ and $V_{lb}$ respectively;
they are equal to
\[ V_{sb} (\pv , \pv')\ =\ \fr{\al \pi}{m^2}\ \left(\
\sisi -\ \fr{1}{\qv^2}\ ( \so \qv )\ ( \sd \qv )\ \right)\ \; ,  \]
\[ V_{lb}( \pv, \pv')\ =\ - 4 \fr{\al \pi}{m^2}\
 \left(\ -\ \fr{(\qv \pv)^2}{q^4}\ +\ \fr{p^2}{q^2}\ \right)\
 , \; \; \; \qv \equiv \pv'-\pv \; \; \; .  \]
As the values to calculate are average values over S-state,
 $V_{lb}$ may be replaced by any function $V_{lb}'$ such that
\[ \int\ V_{lb} ( \pv , \pv' )\ d \vartheta_{\pv \pv'}\ \;
=\ \; \int\ V'_{lb} ( \pv , \pv' )\ d \vartheta_{\pv \pv'}\ \; ; \]
It is convenient to choose
\[ V_{lb}'( \pv, \pv')\ =\ - 4 \fr{\al \pi}{m^2}\
\left(\ - \fr{1}{2}\ +\ \fr{1}{2}\
\left(\ \fr{p'^2}{q^2}\ +\ \fr{p^2}{q^2}\ \right)\ \right)\
\; . \]
For separate parts of
$ E^S_{MM},\ E^S_{MCM}\ $ one obtains
\[ <\ V_{lb}' S_0 R V_{sb}\ +\
 V_{sb} S_0 R V_{lb}'\ >_{(6)}\ =\ \left(\
-\ \fr{1}{3}\ \ln{\al}\ -\ \fr{1}{6}\ \right)\ m \al^6 \; ,  \]
\[ <\ V_{sb} S_0 R V_{sb}\ >_{(6)}\ =\ \left(\
-\ \fr{1}{24}\ \ln{\al}\ -\ \fr{7}{96}\ \right)\ m \al^6 \; ,  \]
\[ <\ V_{lb}' S_0 V_0 S_0 R V_{sb}\ +\
 V_{sb} S_0 V_0 S_0 R V_{lb}'\ >_{(6)}\ =\ \left(\
-\ \fr{1}{6}\ +\ \fr{\pi^2}{18}\ \right)\ m \al^6  \; ,  \]
\[ <\ V_{sb} S_0 V_0 S_0 R V_{sb}\ >_{(6)}\ =\ \left(\
-\ \fr{5}{48}\ \ln{\al}\ +\ \fr{1}{96}\ \right)\ m \al^6  \; ,  \]
which results in
\[ E^S_{MM} =\ \left(\
-\ \fr{3}{8}\ \ln{\al}\ -\ \fr{23}{96}\ \right)\ m \al^6 \; ,  \]
\[ E^S_{MCM} =\ \left(\
-\ \fr{5}{48}\ \ln{\al}\ -\ \fr{5}{32}\ +\ \fr{\pi^2}{18}\
\right)\ m \al^6 \; .  \]

Now the calculation of the "soft" contributions is completed. Their sum is
\[ E^S\ =\ \left(\
-\ \fr{1}{6}\ \ln{\al}\ +\ \fr{1393}{1728}\
\right)\ m \al^6\ \; \approx\ \;
\left(\
-\ \fr{1}{6}\ \ln{\al}\ +\ 0.8061\
\right)\ m \al^6\ \; .  \]

To check the formalism and the calculation technique used, they were applied
to calculation of the recoil
contribution \mbox{$ \sim \al^6 m^2/M $} to the ground state
hyperfine splitting of the hydrogen. It is well known that for this
particular purpose the two-particle problem reduces effectively to
the problem of motion in external field, and the value desired may be easily
obtained by means of coordinate-space calculations, using known
solutions of the Dirac equation in the Coulomb field. The correction to
the hyperfine splitting thus obtained is equal to
\mbox{$ 4 \al^6 m^2 /M $}.
The same result was found using momenta representation by the method
described in the present paper. To simplify corresponding
calculations, diagram technique rules were modified, because in
the relevant order one can consider the magnetic field as permanent, and so
the magnetic photon exchange may be assumed to be instantaneous.
All the "hard" contributions cancel each other; the soft
contributions of the desired order
 arise only from graphs having one magnetic photon, and their
calculation coincides mainly with the calculation of corresponding
contributions for positronium.

\vspace{1.5cm}

5. THE "HARD" CONTRIBUTIONS.

\vspace{0.5cm}

For the calculation of the "hard" contribution to \mbox{$ \De \nu_{rec}\ $},
i.e. the value \mbox{$ <W-W_0>_p\ $}, covariant graphs are used, as
explained in section 3.
In covariant gauge there are only 4 essentially different two-loop
"recoil" graphs, whereas in the Coulomb gauge there are 24 ones (regarding
magnetic and Coulomb quanta as depicted by different lines); here term
"essentially different" mean that these graphs cannot be obtained
from each other by transposition of two particles and/or time reversal.

The hard contribution to \mbox{$ \De \nu_{rec}\ $} was evaluated in two ways.
The first way is to use the Coulomb gauge and to evaluate contributions
of the 24 graphs separately. The second one is to calculate
\mbox{$ <W-W_0>_p\ $} as a whole using the Feynman gauge; in this
case  the $W'$ is sum of contributions of only 4 graphs. Besides that,
these contributions have simpler form than those of the Coulomb gauge (strictly
speaking, the "simpler form" refers to simpler form of the program
calculating these contributions).

The result for the "hard" contribution is
 $-0.424(6) m \al^6\ $ for the "separate" and
$-0.426(6) m \al^6\ $  for the "united" calculation.
Adding the "soft" contribution one obtains
\bq\label{DErec1}
\De \nu_{rec}\ =\ m \al^6\ \left(\ - \fr{1}{6} \ln{\al}\ +\
0.382(6)\ \right)   \ee
and
\bq\label{DErec2}
\De \nu_{rec}\
=\ m \al^6\ \left(\ - \fr{1}{6} \ln{\al}\ +\ 0.380(6)\ \right) \ee
for the "separate" and the "united" calculation methods respectively.
These results
are in perfect agreement with those found in \cite{Pa,Elh,AS}. 

The error estimation quoted in (\ref{DErec1},\ref{DErec2}) is
arbitrary in some extent, being subject of handwork; the reason of
this is that method used in the program for calculation of
\mbox{$ W'\ $} is numerically unstable, and the error quoted arises
from this unstability (and not from the error of numerical integration).
The direct way of the error estimation is to compare the results of
"separate" and "united" calculations; difference between them is
indeed within the estimation quoted above, as well as their differences
from the results of works \cite{Pa,Elh,AS}.  

Results for contributions of separate graphs are quoted in tables 1a,1b,2
together with the revised results of paper \cite{AS},
which agree with the results of the present paper. Errors are not given
in the tables because they are not well defined; in fact, error for
any contribution is \mbox{$ \sim 1\%{} $}.

 Notations for the graphs are following. If one makes no difference
between magnetic and Coulomb lines, there exist 4 essentially
different two-loop recoil graphs, denoted as 1,2,3,4 at fig. 5. To denote
magnetic and Coulomb lines, indices "M" and "C" are written in the
sequence the lines are attached to the lower fermion line from left
to right. 
Values quoted in the tables refer to "essentially different"
graphs only, including in each quoted value contributions of all
graphs of the similar form.

Having been averaged over the full spin direction, $W'$ depends on
three variables; convenient choice for these variables are absolute
values of the loop momenta, referred as
\mbox{$ p_1,\ p_2,\ q\ $}. For the "united" method of the calculation it is
convenient to symmetrize
$W'$ over permutations of \mbox{$ p_1,\ p_2,\ q\ $},
and to take into account only contributions of "essentially different" graphs,
taking some of them multiplied by 2 or 4 because of
transposition of the two particles and/or time reversal. 

Let us describe briefly the checking of the calculation of the "hard"
contributions. The first method
of the checking is to compare  \mbox{$ <W-W_0 >_p $} found in
the two ways, the "separate" and the "united" (and intermediate comparison of
\mbox{$ W'(p_1, p_2, q)\  $} obtained in the two ways, at several
points in \mbox{$ p_1, p_2, q $}).

Second, contributions to $W_0'$ were found by a relatively short analytic
calculation (except contributions associated with sum of graphs 1-CMM
and 2-CMM),
whereas contributions to $W'$ are results of the numerical computing.
An error in
$W'$ or $W_0'$ leads usually to divergency of the numerical
integration; clearly search for such errors is more convenient for
the "separate" calculation.

Third, the results of the "separate" calculation are compared with
the results of \cite{AS} in which \mbox{$\De \nu_{rec} $}
was calculated using
Bethe-Salpeter formalism and the Coulomb gauge. The correction to the
energy can
be written as an expansion in irreducible covariant graphs (except 
the tree graph  with Coulomb photon). Numerical value of any term of this
expansion is clearly independent of formalism used, e.g. the formalism of
the present paper or that of \cite{AS}. So most of contributions
found in the present
paper should be equal to corresponding contributions found in \cite{AS}. The
other terms can be divided into groups, and total contribution of each
group should also be equal to corresponding contribution of \cite{AS}.
Besides that, to order under the consideration the contribution of graph  1-MCM
should be equal separately to corresponding contribution of \cite{AS}.

Results for "hard" contributions of separate graphs are given in
tables 1a,1b.
Part of them (namely, those quoted in table 1b)) are not uniquely defined as
they depend on the choice of $W_0$. On the other hand, contribution of every
term of the expansion in irreducible graphs, mentioned above, does not
depend on formalism used, e.g. on the choice of $W_0$. Some terms of
this expansion are trivial, i.e. they contain only "hard" parts, and
every of them consist of contribution of just one covariant graph;
corresponding results are quoted in table 1a. The other
("non-trivial") terms of this expansion involve also "soft" contributions, and
their "hard" parts may consist of contributions of several covariant graphs;
their values are given in table 2. Tables 1a,2 also contain
corresponding results of \cite{AS}.
 All the results of \cite{AS} are in good agreement with those of 
the present work.

Contributions of graphs 3-MMC and 4-MMC were computed together, because
these graphs contain noncovariant graphs shown in fig. 3(a) and 3(b),
contributions of which
to $V_{30}$ cancel each other, which simplified the calculation.
Contributions of graphs 1-CMM and 2-CMM were also evaluated together due
to the similar reason.

In table 2 the value $E_{1CCC}^H$, for instance, denotes the "hard" part of
the contribution of graph  1-CCC, i.e. one of values quoted in table 1b.
The value $E_M$ stands for
\[ E_{3CCM}^H +E_{2CMC}^H +E_{1MCC}^H +E_{1CMC}^H
+ E^S_{LM}+E^S_M+E_P^S\  , \]
and the $E_M'$ denotes
\[ (ct0+tc0)+ctc0+ (ctx+tcx)+(cct0+tcc0)+ (ctcy+ccty)+tccz+
\De E_{MP}^{hfs} (\delta K_0 \cdot T + T \cdot \delta K_0 )
+ \De E_d^{hfs}\ \; \; \;   \]
(using notations of paper \cite{AS}).

\newpage

\newcommand{\rrr}{\rule[-0.3cm]{0cm}{0.9cm}}
\newcommand{\rr}{\rule[-0.2cm]{2.0cm}{0cm}}
\newcommand{\rrrr}{\rule[-0.2cm]{1.0cm}{0cm}}

\begin{center}
Table 1a. \\ "Trivial" (i.e. containing only the "hard" part)
 contributions to $\De \nu_{rec}$ \\
 (in units \mbox{$ m \al^6\ $}); the results of \cite{AS}
 are also quoted.

\vspace*{0.5 cm}

{\normalsize
   \begin{tabular} {cccc} \hline
     Graph \rrr & \rrrr contribution to $\De \nu_{rec}$  &
     \rrrr  &
 \rrrr contribution to  $\De \nu_{rec}$ \\
  &   &  (notations of \cite{AS})  &  (obtained in \cite{AS}) \\  \hline
   4-CCC \rrr &  -0.0039 &  cccx &  -0.0039  \\
   4-CMC \rrr &   0.0042 &  ctcx &   0.0043  \\
   4-MCC \rrr &  -0.0486 &  cctx + tccx &  -0.0489  \\
   4-MCM \rrr &  -0.0230 &  tctx &  -0.0230  \\
   1-MMM \rrr &   0.0694 &  ttt0 &  0.0694  \\
   2-MMM \rrr &  -0.0011 &  ttty &  -0.0011  \\
   3-MMM \rrr &  -0.0012 &  tttz &  -0.0011  \\
   4-MMM \rrr &   0.0042 &  tttx &  0.0041  \\
   3-CCC \rrr &   0.0064 &  cccz &  0.0063  \\
   3-MCC \rrr &   0.0268 &  ctcz+cctz &  0.0283  \\
   3-CMM \rrr &   0.0530 &  ttcz+tctz &  0.0534  \\
   2-CCC \rrr &  -0.0184 &  cccy+ccx &  -0.0186  \\
   2-MCC \rrr &  -0.0661 &  tccy &  -0.0681  \\
   2-MCM \rrr &   0.0552 &  tcty+ttcy &  0.0558  \\
 \hline  \end{tabular} }
\end{center}

\newpage
\begin{center}
Table 1b. \\ "Hard" parts of "non-trivial"
 contributions to $\De \nu_{rec}$  (in units \mbox{$ m \al^6\ $}).
{\normalsize
   \begin{tabular} {ccccccc} \hline
     Graph \rrr & \rr \rr \rr \rr  & \rr  \\ \hline
   1-CCC \rrr &   & -0.0094  \\
   1-MCM \rrr &   & -0.0745  \\
   1-CMM + 2-CMM \rrr &   & -0.5112  \\
   4-MMC + 3-MMC \rrr &   & -0.0209  \\
   3-CCM \rrr &   & -0.0104  \\
   2-CMC \rrr &   & -0.0245  \\
   1-MCC \rrr &   & 0.1795  \\
   1-CMC \rrr &   & -0.0092  \\
 \hline  \end{tabular} }

\vspace*{1.0 cm}
Table 2. \\ "Non-trivial"  contributions to $\De \nu_{rec}$ 
 (in units \mbox{$ m \al^6\ $}), \\
 compared with the results of \cite{AS}.

\vspace*{0.5 cm}

{\normalsize
   \begin{tabular} {ccccc} \hline
value quoted&value quoted&  coefficient &      & constant \\
            & (notations of \cite{AS}) & at $\ln{\al}$& constant  &
(obtained in \cite{AS})  \\
\hline
 $E^H_{4MMC+3MMC}$ \rrr& (cttx+ttcx)+cttz & 0 & -0.0209 & -0.0209 \\
\hline
\rrr $E^H_{1CCC}+E_C^S$& cc0+ccc0 & $ - 1/48$& -0.0146 & -0.0148 \\
\hline
\rrr $E^H_{1MCM} +E^S_{MCM}$& tct0 & $ - 5/48 $& 0.3176 & -0.3138 \\
\hline
$E^H_{1CMM+2CMM} +E^S_{MM}$
 & ctt0+ttc0+ctty + & $ - 3/8 $& -0.751 &-0.749 \\
    & + tt0+ttx & & &  \\
\hline
 \rrr $ E_{M} $& $  E_{M}'$ & 1/3 & 0.427 & 0.423 \\
\hline
 \end{tabular} }
\end{center}

\vspace{0.5cm}

 6. SUMMARY AND DISCUSSION.

\vspace{0.5cm}

In the present paper the "recoil" contribution of order $ m \al^6$
to the hyperfine splitting of positronium ground state has been found.
The calculation was performed using noncovariant formulation of QED
perturbation theory. The result equals 
\[
\De \nu_{rec}\
=\ m \al^6\ \left(\ - \fr{1}{6} \ln{\al}\ +\ 0.381(6)\ \right)
\;  \]
and is in perfect agreement with the results of \cite{Pa,Elh,AS}.  
Let us note that
methods used in \cite{Pa,Elh} and in the present work are
similar, whereas that of \cite{AS} is completely different. 
So \mbox{$ \De \nu_{rec}\ $} may be considered to be 
firmly established. However agreement with the experimental data is poor.
Combining nonlogarithmic part of the recoil contribution with all other
contributions of the relevant order (they also may be considered to be
reliable), one obtains the complete nonlogarithmic contribution
 $\sim m \al^6$ to $\De \nu$:
\bq\label{denu6}
 m \al^6\ (\ -0.3928\ )\ =\ -\ 7.33\ MHz\   \ee
 ((\ref{denu6}) was obtained using for \mbox{$ \De \nu_{rec}$}
 the result of \cite{Elh}, as the most accurate).
 The complete theoretical result up to the order \mbox{$m \al^6 $}
is sum of (\ref{denu6}) and (\ref{denu456}), and equals 
\[ \De \nu_{th}\ =\ 203\ 392.96\  MHz\ \; .   \]
It differs from the experimental result (\ref{exp2}) by 5
standard deviations.

The leading term of the next order in $\al$, i.e. of order
\mbox{$ \sim m \al^7\ \ln^2 {\al}\ $}, was found in \cite{Kar} and
equals 
\[  -\ \fr{7}{8 \pi}\ m \al^7 \ln^2 {\al}\
=\ -0.92 MHz\  .   \]
Taking this correction into account reduces the difference between
experimental and theoretical results to 4 standard deviations. One cannot
reject possibility that this difference may be explained by contributions
\mbox{$ \sim m \al^7\ \ln {\al},\ \sim m \al^7\  )\ $}).

I am grateful to
I.B.Khriplovich, A.I.Milstein, and A.S. Yelkhovsky for useful
conversations and interest to this work. I am also grateful to
A.S.Yelkhovsky for informing me about the result of \cite{Elh} prior to
publication, and G.S.Adkins for useful communication.  
I am also deeply grateful to D.Yu.Ivanov for his
co-operation in the early stage of the work.

\newcommand{\vpl}[5]{ \put(#1,#2){\begin{picture}(80,20) 
 \multiput(0,#3)(0,#4){#5}{\oval(#3,#3)[l]}
 \multiput(0,0)(0,#4){#5}{\oval(#3,#3)[r]}  \end{picture}}  }

\newcommand{\vplo}[6]{ \put(#1,#2){\begin{picture}(80,20) 
 \multiput(0,#3)(0,#4){#5}{\oval(#6,#3)[l]}
 \multiput(0,0)(0,#4){#5}{\oval(#6,#3)[r]}  \end{picture}}  }

\newcommand{\ulpl}[6]{ \put(#1,#2){\begin{picture}(80,20) 
 \multiput(0,0)(#3,#4){#6}{\oval(#5,#4)[tr]}
 \multiput(0,#4)(#3,#4){#6}{\oval(#5,#4)[bl]}  \end{picture}}  }

\newcommand{\urpl}[5]{\put(#1,#2){\begin{picture}(80,20) 
 \multiput(0,0)(#3,#4){#5}{\oval(#3,#4)[tl]}
 \multiput(0,#4)(#3,#4){#5}{\oval(#3,#4)[br]}  \end{picture}}  }

\newcommand{\varr}[2]{\put(#1,#2){\vector(0,1){0}}}
\newcommand{\harr}[2]{\put(#1,#2){\vector(1,0){0}}}

\newcommand{\feb}[4]{  \put(#1,#2){\begin{picture}(80,20) 
     \thicklines
     \put(0,0){\vector(1,0){#3}}
     \put(0,#4){\vector(1,0){#3}}
  \end{picture}} }

\newcommand{\fl}[5]{ \put(#1,#2){\begin{picture}(10,10)  
\thicklines 
 \put(0,0){\line(#3,#4){#5}}  \end{picture}}  }

\newcommand{\cl}[5]{ \put(#1,#2){\begin{picture}(80,20) 
 \multiput(0,0)(0,#3){#5}{\line(0,1){#4}}
  \end{picture}}  }

\newpage

\begin{figure}
\begin{picture}(460,560)
\setlength{\unitlength}{0.16mm}  

     \put(300,1100){\begin{picture}(460,140)
     \setlength{\unitlength}{0.2mm}  
     \feb{0}{0}{300}{100}   
     \urpl{50}{0}{30}{20}{5}
     \urpl{100}{0}{30}{20}{5}
     \put(100,-60){\large Fig. 1 }
     \end{picture}
     }

     \put(20,650)
     {\begin{picture}(460,140)
     \feb{0}{0}{350}{160}   
        \put(0,0)
        {\begin{picture}(20,20)
        \ulpl{250}{0}{-26}{20}{26}{8}
        \varr{107}{120}
        \put(80,110){$\qv$}
        \end{picture} 
        }
        \put(160,0)
        {\begin{picture}(20,20)
        \cl{0}{0}{28}{20}{6}
        \varr{0}{110}
        \put(10,110){$\qv'$}
        \end{picture} 
        }
     \put(20,167){$\pv$}
     \put(80,167){$\kv_1$}
     \put(250,167){$\kv$}
     \put(190,-25){$\pv_1$}

     \put(150,-80){(a)}
     \end{picture}
     }

     \put(550,650)
     {\begin{picture}(460,140)
     \feb{0}{0}{420}{160}   
        \put(0,0)
        {\begin{picture}(20,20)
        \ulpl{360}{0}{-40}{20}{40}{8}
        \varr{140}{120}
        \put(110,110){$\qv$}
        \end{picture} 
        }
        \put(280,0)
        {\begin{picture}(20,20)
        \cl{0}{0}{28}{20}{6}
        \varr{0}{30}
        \put(-30,20){$\qv_2$}
        \end{picture} 
        }
        \put(170,0)
        {\begin{picture}(20,20)
        \cl{0}{0}{28}{20}{6}
        \varr{0}{70}
        \put(-30,60){$\qv_1$}
        \end{picture} 
        }
     \put(30,170){$\pv$}
     \put(100,170){$\kv_2$}
     \put(210,170){$\kv_1$}
     \put(330,170){$\kv$}
     \put(210,-25){$\pv_1$}
     \put(310,-25){$\pv_2$}
     \put(180,-80){(b)}
     \end{picture}
     }

     \put(200,320)
     {\begin{picture}(460,140)
      \setlength{\unitlength}{0.21mm}  
     \feb{0}{0}{350}{120}   
     \ulpl{300}{0}{-30}{15}{30}{8}
     \cl{120}{0}{26}{16}{5}
     \cl{190}{0}{26}{16}{5}
     \cl{260}{0}{26}{16}{5}
     \put(150,-60){(c)}
     \end{picture}
     }

     \put(460,120){\large Fig.2} 

\end{picture}
\end{figure}

\newpage 
\clearpage 
\newpage  

\begin{figure}
\begin{picture}(460,540)

  \put(0,420){\begin{picture}(460,140)
  \setlength{\unitlength}{0.09mm}  

        \put(350,600)
        {\begin{picture}(460,140)
        \fl{160}{0}{-1}{0}{400}
        \fl{-160}{160}{1}{0}{400}
        \cl{0}{0}{28}{20}{6}
        \ulpl{134}{0}{-52}{40}{52}{5}
        \ulpl{74}{-40}{-52}{40}{52}{5}
        \fl{160}{0}{-3}{-2}{60}
        \fl{-160}{160}{3}{2}{60}
        \fl{100}{-40}{1}{0}{200}
        \fl{-100}{200}{-1}{0}{200}

        \put(0,-100){(a)}
        \end{picture}
        }
        
        \put(750,700){\large $+$}

       \put(1250,600)
       {\begin{picture}(460,140)
        \fl{160}{0}{-1}{0}{450}
        \fl{-160}{160}{1}{0}{450}
        \cl{0}{0}{28}{20}{6}
        \ulpl{80}{-40}{-40}{48}{40}{5}
        \ulpl{140}{0}{-40}{20}{40}{8}
        \fl{160}{0}{-3}{-2}{60}
        \fl{-160}{160}{3}{2}{60}
        \fl{-100}{200}{-1}{0}{150}
        \fl{100}{-40}{1}{0}{150}
        \put(0,-100){(b)}
        \end{picture}
        }
 
        \put(1800,700){\large $\approx$}
        \put(300,300){\large $\approx$}
         
        \put(1150,120)
        {\begin{picture}(460,140)
         \setlength{\unitlength}{0.16mm}

        \fl{-240}{0}{1}{0}{320}
        \fl{-240}{140}{1}{0}{320}
        \ulpl{-5}{0}{-10}{20}{10}{4}
        \ulpl{-50}{80}{-20}{10}{20}{6}
        \ulpl{-10}{0}{-20}{10}{20}{6}
        \ulpl{-125}{60}{-10}{20}{10}{4}
        \put(0,0){\circle*{10}}
        \put(-160,140){\circle*{10}}
        \cl{-80}{0}{30}{20}{5}
        \put(-80,-40){(c)}
        \end{picture}
        }

   \end{picture} 
   }

  \put(0,140){\begin{picture}(460,140)
  \setlength{\unitlength}{0.2mm}

        \put(0,400)
        {
        \begin{picture}(460,140)
        \setlength{\unitlength}{0.14mm}
        \fl{60}{20}{-1}{0}{120}
        \fl{0}{0}{1}{0}{120}
        \fl{0}{0}{3}{1}{60}
        \ulpl{115}{-100}{-10}{20}{10}{6}
        \urpl{-45}{-100}{10}{20}{5}
        \put(0,-200){(d)}
        \end{picture}
        }
        
        \put(180,380){\large $+$}
 
        \put(300,400)
        {\begin{picture}(460,140)
        \setlength{\unitlength}{0.14mm}
        \fl{57}{19}{-1}{0}{120}
        \fl{0}{0}{1}{0}{120}
        \fl{0}{0}{3}{1}{57}
        \ulpl{110}{-120}{-20}{20}{20}{6}
        \urpl{-75}{-121}{20}{20}{7}
        \put(0,-200){(e)}
        \end{picture}
        }

        \put(480,380){\large $\approx$}

        \put(700,400)
        {\begin{picture}(460,140)
        \fl{-100}{0}{1}{0}{200}
        \put(0,0){\circle*{8}}
        \ulpl{55}{-84}{-10}{14}{10}{6}
        \urpl{-55}{-84}{10}{14}{6}
        \put(0,-150){(f)}
        \end{picture}
        }

   \end{picture} 
   }

  \put(0,0){\begin{picture}(460,140)
  \setlength{\unitlength}{0.09mm} 

        \put(350,600)
        {\begin{picture}(460,140)
        \fl{160}{0}{-1}{0}{400}
        \fl{-160}{160}{1}{0}{400}
        \ulpl{134}{0}{-52}{40}{52}{5}
        \ulpl{74}{-40}{-52}{40}{52}{5}
        \fl{160}{0}{-3}{-2}{60}
        \fl{-160}{160}{3}{2}{60}
        \fl{100}{-40}{1}{0}{200}
        \fl{-100}{200}{-1}{0}{200}

        \put(0,-100){(g)}
        \end{picture}
        }
        
        \put(750,700){\large $+$}

       \put(1250,600)
       {\begin{picture}(460,140)
        \fl{160}{0}{-1}{0}{450}
        \fl{-160}{160}{1}{0}{450}
        \ulpl{80}{-40}{-40}{48}{40}{5}
        \ulpl{140}{0}{-40}{20}{40}{8}
        \fl{160}{0}{-3}{-2}{60}
        \fl{-160}{160}{3}{2}{60}
        \fl{-100}{200}{-1}{0}{150}
        \fl{100}{-40}{1}{0}{150}
        \put(0,-100){(h)}
        \end{picture}
        }
   
        \put(1800,700){\large $\approx$}
        \put(300,300){\large $\approx$}
         
        \put(1150,150)
        {\begin{picture}(460,140)
         \setlength{\unitlength}{0.16mm}

        \fl{-240}{0}{1}{0}{320}
        \fl{-240}{140}{1}{0}{320}
        \ulpl{-5}{0}{-10}{20}{10}{4}
        \ulpl{-50}{80}{-20}{10}{20}{6}
        \ulpl{-10}{0}{-20}{10}{20}{6}
        \ulpl{-125}{60}{-10}{20}{10}{4}
        \put(0,0){\circle*{10}}
        \put(-160,140){\circle*{10}}
        \put(-80,-40){(i)}
        \end{picture}
        }
       
        \put(650,-100){\large Fig. 3}

   \end{picture} 
   }

\end{picture}
\end{figure}

\newpage
\clearpage 
\newpage  

\begin{figure}
\begin{picture}(460,540)  
\setlength{\unitlength}{0.14mm}

\put(0,700){\begin{picture}(460,140)

   \put(30,400)
   {\begin{picture}(460,140)
   \feb{0}{0}{350}{160}   
        \put(0,0)
        {\begin{picture}(20,20)
        \ulpl{250}{0}{-26}{20}{26}{8}
        \varr{107}{125}
        \put(80,110){$\qv$}
        \end{picture} 
        }
        \put(190,0)
        {\begin{picture}(20,20)
        \cl{0}{0}{28}{20}{6}
        \varr{0}{110}
        \put(10,110){$\qv'$}
        \end{picture} 
        }
   \put(20,170){$\pv$}
   \put(250,170){$\pv'$}

   \put(150,-70){(A)}
   \end{picture}
   }

   \put(650,400)
   {\begin{picture}(460,140)
   \feb{0}{0}{350}{160}   
        \put(90,0)
        {\begin{picture}(20,20)
        \vplo{0}{10}{20}{40}{4}{10}
        \varr{5}{90}
        \put(10,80){$\qv$}
        \end{picture} 
        }
        \put(230,0)
        {\begin{picture}(20,20)
        \cl{0}{0}{28}{20}{6}
        \varr{0}{80}
        \put(10,70){$\qv'$}
        \end{picture} 
        }
   \put(40,170){$\pv$}
   \put(310,170){$\pv'$}

   \put(150,-70){(B)}

   \end{picture}
   }

   \put(500,230){\large Fig. 4}

\end{picture}
}

\put(0,-80){\begin{picture}(460,140)
\setlength{\unitlength}{0.2mm}

   \put(50,400)
   {
   \begin{picture}(460,140)
   \feb{0}{0}{250}{100}   
   \put(60,0){\line(0,1){100}}
   \put(130,0){\line(0,1){100}}
   \put(200,0){\line(0,1){100}}
   \put(120,-40){(1)}
   \end{picture}
   }

   \put(450,400)
   {
   \begin{picture}(460,140)
   \feb{0}{0}{250}{100}   
   \put(50,0){\line(0,1){100}}
   \put(120,0){\line(1,1){100}}
   \put(120,100){\line(1,-1){100}}
   \put(120,-40){(2)}
   \end{picture}
   }

   \put(50,190)
   {\begin{picture}(460,140)
   \feb{0}{0}{250}{100}   
   \put(100,0){\line(0,1){100}}
   \put(150,0){\line(0,1){100}}
   \put(50,100){\line(3,-2){150}}
   \put(120,-40){(3)}
   \end{picture}
   }

   \put(450,190)
   {\begin{picture}(460,140)
   \feb{0}{0}{250}{100}   
   \put(125,0){\line(0,1){100}}
   \put(50,0){\line(3,2){150}}
   \put(50,100){\line(3,-2){150}}
   \put(120,-40){(4)}
   \end{picture}
   }

   \put(370,50){\large Fig. 5}

\end{picture}  
}

\end{picture}
\end{figure}

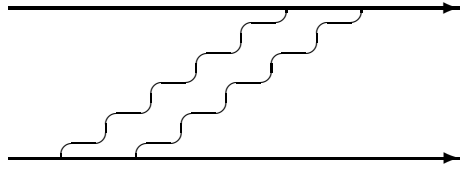
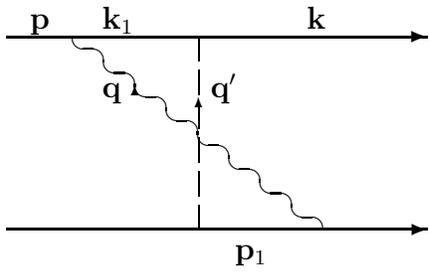
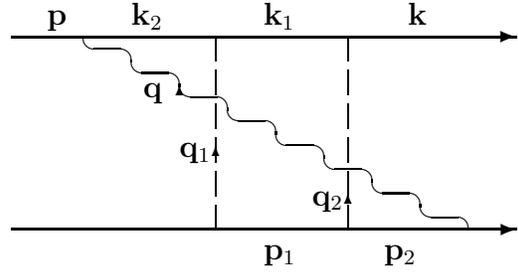
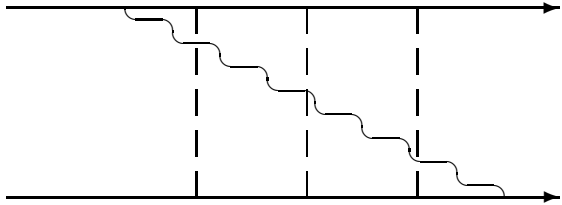
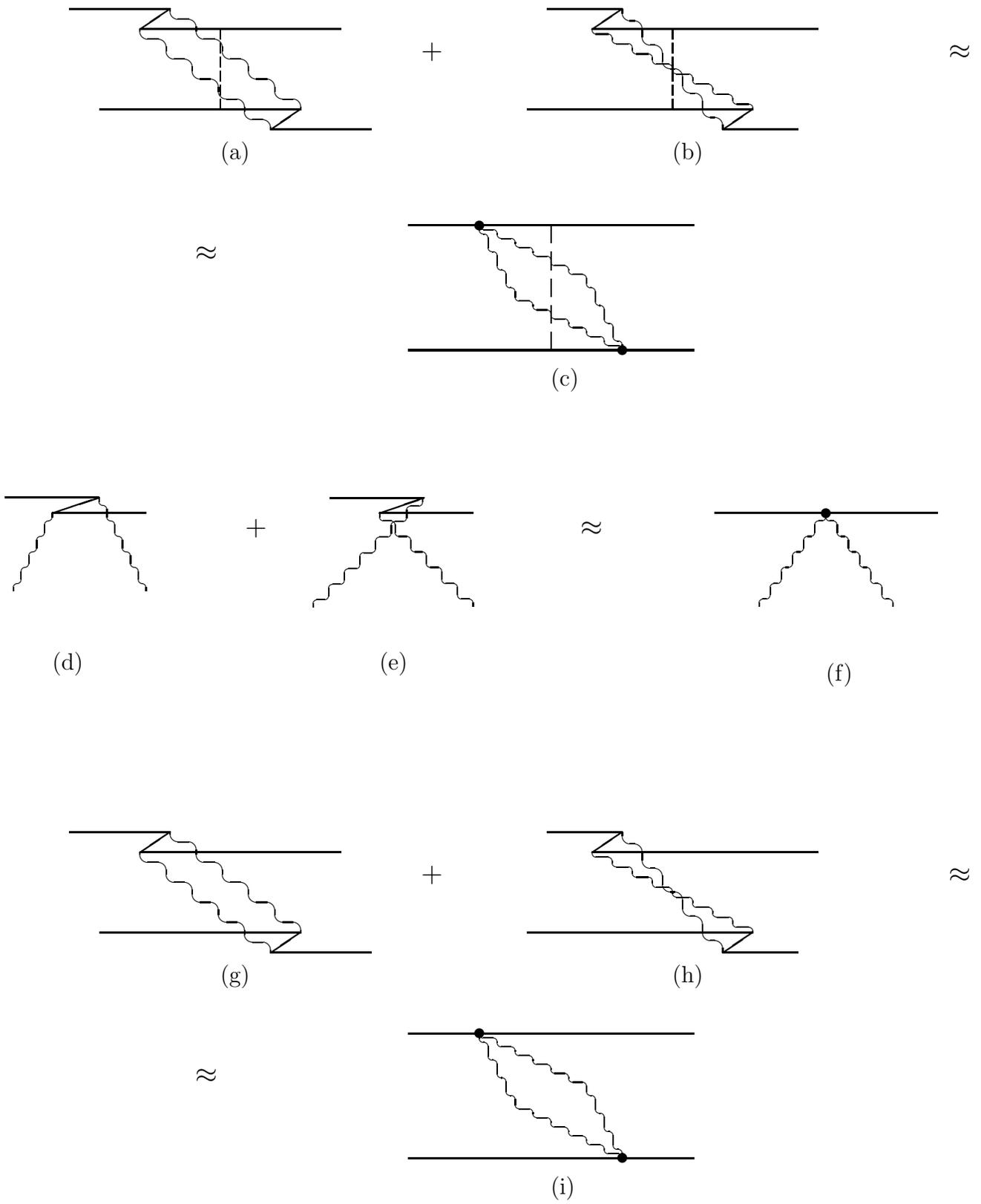
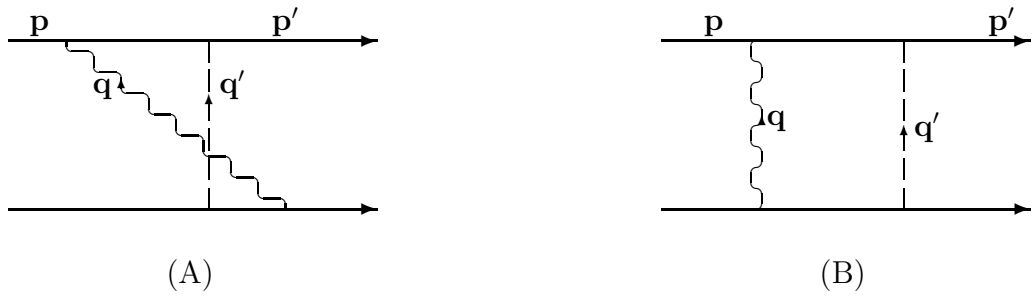
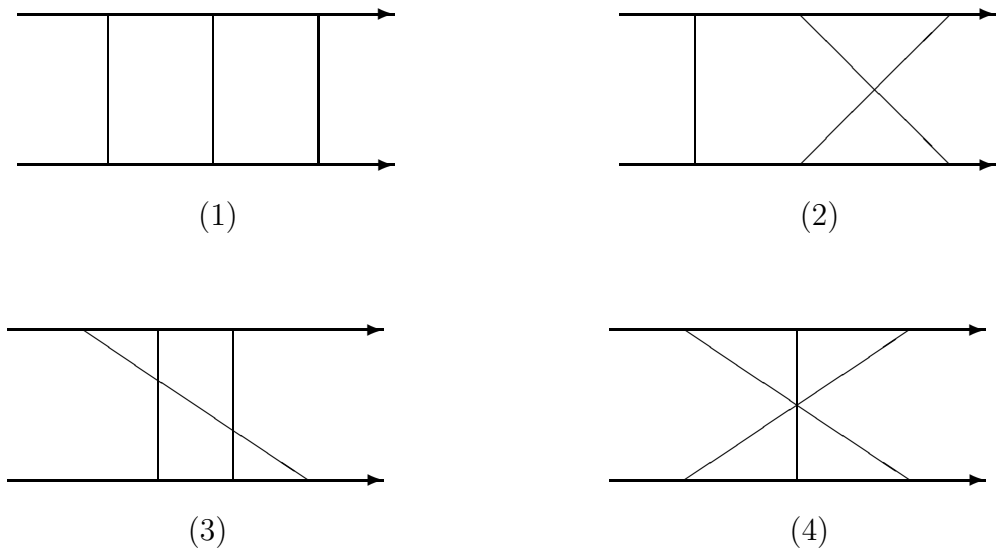
\end{document}